\documentclass{aa}  

\usepackage{graphicx}
\usepackage{txfonts}
\usepackage[colorlinks=true, citecolor=blue]{hyperref}
\usepackage{array}
\usepackage{tabularx}
\usepackage{amsmath}
\usepackage{multirow}
\usepackage{natbib}
\usepackage{tablefootnote}
\usepackage{verbatim}
\bibpunct{(}{)}{;}{a}{}{,} 

\usepackage{comment}
\usepackage{color}
\usepackage[normalem]{ulem}

\newcommand{\dgs}[1]{{\color{red}{#1}}}

\newcommand{\kernelp}{W(\vert {\bf r'}-{\bf r}\vert,h)}

\newcommand{\kernelsha}{W(\vert {\bf r}_b-{\bf r}_a\vert,h_a)}
\newcommand{\kernel}{W_{ab}(h_a)}

\makeatletter 
\renewcommand*\aa@pageof{, page \thepage{} of \pageref*{LastPage}}
\makeatother  

\begin{document}

\title{Conservative, density-based  smoothed
particle hydrodynamics with improved partition of the unity and better estimation of gradients}   
   \author{Domingo García-Senz\inst{1}
          \and
          Rubén M. Cabezón\inst{2}
          \and
          Jose A. Escartín\inst{3}
          }

   \institute{Departament de Física. Universitat Politècnica de Catalunya (UPC). Avinguda Eduard Maristany 16, E-08019 Barcelona (Spain)\\
              \email{domingo.garcia@upc.edu}
         \and
             Center for Scientific Computing - sciCORE. Universität Basel, Klingelbergstrasse 61, 4056, Basel  (Switzerland)\\
             \email{ruben.cabezon@unibas.ch}
             \and
             Euclid Science Data Center - Max Planck Institute for Extraterrestrial Physics, Giessenbachstrasse 1, 85748 Garching, (Germany)\\
             \email{jescarti@mpe.mpg.de}
             }

   \date{XX-07-2021}

 
  \abstract
   {The accurate evaluation of gradients is a cornerstone of the smoothed particle hydrodynamics (SPH) technique. Using an integral approach to estimating gradients has been proven to substantially enhance its accuracy, retaining the Lagrangian structure of SPH equations and remaining fully conservative. However, in practice, it is difficult to ensure that the Lagrangian formulation is entirely consistent with regard to the exact partition of the unity.}
   {In this paper, we focus our study on the connection between the choice of the volume elements (VEs) in the SPH summations as well as the accuracy in the gradient estimation within the integral approach scheme (ISPH). We propose a new variant of VEs to improve the partition of the unity that is fully compatible with the Lagrangian formulation of SPH, including grad-h corrections.}
   {Using analytic considerations, simple static toy models in 1D, and a set of full 3D test cases, we show that any improvement in the partition of the unity also leads to a better calculation of gradients when the integral approach is used jointly. Additionally, we propose an easy-to-implement modification of the ISPH scheme, which  makes it more flexible and better suited to handling sharp density contrasts.}
   {The ISPH code that is built with the proposed scheme has been validated with a good number of standard tests, some of them involving contact discontinuities. The performance of the code was shown to be excellent in all of these tests, consistently demonstrating that an improvement in the partition of the unity is not detrimental to the optimal conservation of energy, momentum, and entropy that is typical of Lagrangian schemes.}
   {We successfully built a new ISPH scheme on a Lagrangian basis, which is fully conservative, and compatible with self-consistent grad-h terms and an improved partition of the unity. The ensuing code is able to successfully cope with the tensile instability and has been validated with a number of hydrodynamic tests with good results.}

   \keywords{ Methods: numerical  -hydrodynamics}
               
\titlerunning{Conservative SPH with improved partition of the unity and gradients}
\maketitle
%

\section{Introduction}
The smoothed particle hydrodynamics (SPH) is a firmly settled numerical technique that is capable of successfully simulating many cutting-edge problems in physics, astrophysics, and engineering. This technique has undergone a sustained enhancement since its original formulation \citet{lucy77}; \citet{gingold77} and it is still evolving at a suitable pace \citep[e.g.,][]{spr10, monaghan12, ros15, Wadsley17, cabezon2017, price18, rosswog2020magma}. A landmark in that evolution concerns the estimation of derivatives and gradients, which can be done using a number of different approaches. 

The standard way of calculating gradients is by directly taking the analytic derivative of the interpolating kernel function, which leads to E0-errors, even in the presence of constant pressure fields in non-uniform particle distributions \citep{age07, zhu15}. One alternative adapts the moving-least squares technique (MLS) to SPH \citep{dilts99} to ensure an exact interpolation of linear functions. In the MLS methods, the SPH kernel is replaced by a new interpolant that is built by combining the original kernel with a linear operator. Such techniques ensure the linear consistency of the interpolations but they also require the solution of a linear system of $d+1$ equations (where $d$~is the dimension) for each particle at each integration step. Alternatively, adding renormalization corrections to both the kernel and the gradient of the kernel have been shown to enhance the accuracy in the calculation of gradients and to reduce the tensile instability \citep{bonet99}. Tensile instability often arises at boundaries in the flow, separating fluid regions of varying densities. The inconsistency in the treatment of gradients induces an nonphysical surface tension of numerical origin, which inhibits fluid mixing and suppresses hydrodynamic instabilities. A variational principle applied jointly with a kernel renormalization was used by \citet{oger07} to reduce the tensile instability and simulate fluid systems with free surfaces. Nevertheless, these MLS and renormalization techniques, in general, do not guarantee the perfect conservation of the whole set of physical laws governing the motion of the fluid, which are at the foundations of the SPH technique. A recent proposal has been the conservative reproducing kernel SPH method (CRKSPH) from \citet{fro17} and although it is not derived from a Lagrangian, it still enforces a perfect linear interpolation and retains the linear momentum and energy conservation properties.

Another alternative way of estimating gradients was devised by \citet{garciasenz2012}. In their proposal, gradients are calculated from an integral expression, so that there is no need to explicitly calculate the analytic derivative of the kernel function. These authors also proved that such an integral approach can be completely compatible with the Lagrangian formulation of the SPH equations, leading to the integral smoothed particle hydrodynamics (ISPH) scheme, named IAD$_0$ in the seminal paper by \citet{garciasenz2012}. It has been shown that the ISPH formulation has the same conservation properties as the standard Lagrangian-derived SPH \citep{spr02}. We note that a particularly remarkable feature of ISPH is that it effectively reduces the E0-error in the derivatives \citep{garciasenz2012, ros15, val16}.

In this work we dig further into the conditions that the ISPH scheme should meet in order to improve the calculation of gradients, which can approach greater precision for linear functions. We found that these conditions are connected with another basic SPH requirement, namely: the correct partition of unit volume. By using one-dimensional numerical experiments, we clarify the link between these two basic properties, namely: any enhancement in the partition of unity leads to a better gradient estimation within the ISPH framework. The results of these 1D toy models are confirmed by detailed 3D hydrodynamic simulations of explosions, collisions, and instabilities. 

Here, we propose and discuss a new type of volume elements (VEs) that lead to a better partition of the unity. The ISPH equations of density, movement, and energy are consequently re-formulated, within the Lagrangian framework, so that they become fully compatible with the particular choice of the VEs. The resulting scheme is not only fully conservative, but it also enhances the density estimation and the gradient of linear functions with practically no computational overload. 

Thus, the present work is a natural extension of the discussion in \citet{cabezon2017} (hereafter Paper I), where the code SPHYNX was introduced and verified. Here, we focus  on modifying the proposal in Paper I, so that the new SPH scheme (Appendix \ref{sec:Appendix}) incorporates an implicit treatment of the VEs, which is now fully compatible with the Lagrangian derivation of SPH equations, with a consistent treatment of grad-h terms. A second improvement is the use of a novel self-adaptive scheme to surgically reduce the tensile instability in regions with large density gradients. Unfortunately, suppressing that instability  requires us to abandon the Lagrangian formulation \citep{rea10, Wadsley17, wissing20}. In our proposal, the departure from such a formulation is minimal and is limited to regions hosting significant density contrasts. Additionally, we show that our formulation leads to a better behavior in terms of entropy in shocks than other schemes that suppress the tensile instability.

In Section~\ref{sec:isph}, we review the main features of the ISPH scheme. We discuss the choice of the generalized volume elements used to compute the summations in Section~\ref{sec:volume-elements}. The link between the performance of the ISPH calculation of gradients and the adequate choice of the VEs is highlighted in Section~\ref{sec:static}. Section~\ref{sec:isph_equations0} presents the resulting ISPH equations, while in Section~\ref{sec:dynamic}, we apply our code to several standard tests calculated in three dimensions and we analyze the results in the context of the VEs choice. A summary of our findings and prospects for future works is given in the conclusions section.

\section{ISPH formulation}
\label{sec:isph}
The classical way of evaluating gradients in SPH takes the multidimensional derivative of a function $f$ as:

\begin{equation}
{\bf \nabla}f = \int_{V} f({\bf r'})~\nabla \kernelp~dr'^3\,,
\label{Interpolation1}
\end{equation}

\noindent
where $\kernelp$ is commonly a Dirac $\delta$-like function named interpolating kernel, which is continuous and derivable, and $h$~is the smoothing length. In the classical SPH formulation, the gradient of the function is estimated by: {\bf 1)} approaching the integral of Eq.~(\ref{Interpolation1}) by summations; {\bf 2)} taking the analytic derivative of the kernel; and {\bf 3)} assuming that the volume element $dr'^3$ is adequately represented by $m/\rho$. Specifically\footnote{Admittedly, this is not the most used implementation of the derivative in SPH. The most commonly used version is a variation involving differences of the derived function. Nevertheless, it is formally the same as Eq.~(\ref{Interpolation2}).},

\begin{equation}
\left<{\bf\nabla}f\right>_a = \sum_b \frac{m_b}{\rho_b}f_b\nabla \kernelsha\,,
\label{Interpolation2}
\end{equation}

\noindent where $a$ and $b$ refer to neighbouring particles with masses $m_a$ and $m_b$, respectively. Alternatively, in ISPH, the gradient is calculated from an integral approach (IA), which does not require the explicit analytic derivative of $W (\vert{\bf r}-{\bf r'}\vert,h)$. A vector integral $I({\bf r})$ is defined as \citep{garciasenz2012}: 
\begin{equation}
I({\bf {r}})\equiv \int_{V} \left[f({\bf {r'}})-f({\bf {r}})\right] ({\bf {r'}}-{\bf {r}}) \kernelp ~dr'{^3}\,.
\label{int}
\end{equation}

\noindent  The integral $I({\bf {r}})$ can be used to find the gradient of a function $f({\bf {r}})$ in a similar way that the Laplace operator is usually approached from another integral expression in standard SPH  \citep{brookshaw1985, mon05}. The IA interpretation of SPH is the consequence of approaching Eq.~(\ref{int}) with summations, along with approaching the function $f({\bf {r}})$ by a Taylor expansion around the evaluated point,
\begin{equation}
f({\bf r}_b)-f({\bf r}_a)\simeq\nabla {\bf f}_a\cdot({\bf r}_b-{\bf r}_a)\,,
\label{gradient}
\end{equation}

 \noindent with $\nabla{\bf f}_a$~defined below. The RHS in the integral expression in Eq.~(\ref{int}) becomes:

\begin{equation}
\begin{split}
I({\bf r}_a)=
\left[\sum_b\frac{m_b}{\rho_b} f({\bf r}_b)({\bf r}_b-{\bf r}_a) \kernelsha\right]-
\left[f({\bf r}_a) \left<{\bf \Delta r}\right>_a\right]\,,
\end{split}
\label{approxI}
\end{equation}

\noindent
where
\begin{equation}
\left<{\bf \Delta r}\right>_a = \sum_b\frac{m_b}{\rho_b} ({\bf r}_b-{\bf r}_a) \kernelsha\,.
  \label{Deltar}
\end{equation}

Setting  Eqs.~(\ref{gradient}) and (\ref{approxI}) into  Eq.~(\ref{int}) allows us to obtain the gradient $\nabla{\bf f}_a$,
\begin{equation}
\left[
\begin{array}{c}
\partial f/\partial x_1\\
\partial f/\partial x_2\\
\partial f/\partial x_3\\
\end{array}
\right]_a
=
\left[
\begin{array}{ccc}
\tau_{11} & \tau_{12} & \tau_{13} \\
   \tau_{21}&\tau_{22}&\tau_{23} \\
   \tau_{31}&\tau_{32}&\tau_{33}
\end{array}
\right]^{-1}
\left[
\begin{array}{c}
I_1\\
I_2\\
I_3\\
\end{array}
\right]\,,
\label{matrix}
\end{equation}

\noindent where
\begin{equation}
\tau_{ij,a}=\sum_b \frac{m_b}{\rho_b}(x_{i,b}-x_{i,a})(x_{j,b}-x_{j,a})W_{ab}(h_a)\,; i,j=1,3\,,
\label{tauijsph}
\end{equation}

\noindent and where the spatial coordinates are represented by $x$ with sub-indexes $i$ or $j$. Henceforth, we note $\kernel\equiv\kernelsha$ for the sake of clarity.

It was shown in \citet{garciasenz2012} that Eq.~(\ref{matrix}) leads to a perfect linear interpolation. Unfortunately, the price to pay is the loss of the full conservation of linear and angular momentum. The exact conservation of the SPH Euler equations and a perfect linear interpolation can only be retrieved simultaneously when $\left<{\bf \Delta r }\right>_a\rightarrow 0$. We refer to ISPH as the conservative scheme which simply neglects the term $f({\bf r}_a)\left<{\bf \Delta r}\right>_a$ in Eq.~(\ref{approxI}). This is justified because $\left<{\bf \Delta r}\right>$ is, in fact, the indefinite integral of an odd function, which is zero\footnote{Interestingly, this argument has been used for the whole history of SPH to justify that it is a second-order method, which is true in the continuum limit, but not always numerically ensured.}.

On the other hand, the complete integral approach, which takes into account the $f({\bf r}_a)\left<{\bf \Delta r }\right>_a$ term in the RHS of Eq.~(\ref{approxI}), leads to a perfect linear interpolation but is not fully conservative. We refer to it as non-conservative ISPH (ncISPH, hereafter). As commented above, both schemes, ISPH and ncISPH, converge to the same outcome when $\left<{\bf \Delta r }\right>_a\simeq 0$. Having both, a perfect partition of unity and $\left<{\bf \Delta r}\right>_a  = 0$, has been identified for a long time as the main constraint in order to ensure complete linear consistency in SPH \citep{liu06}.

\section{Choice of the volume elements}
\label{sec:volume-elements}

 A common choice with regard to the volume elements in SPH is $V_a=m_a/\rho_a$, which leads to the widely used density equation, 

\begin{equation}
\rho_a=\sum_{b=1}^{n_b}~V_b~\rho_b~ \kernel=\sum_{b=1}^{n_b} m_b \kernel\,,
\label{density}
\end{equation}

\noindent
  which works well provided that the density does not change very much within the kernel range. Nevertheless, the density may appreciably be miscalculated in the presence of shocks and density discontinuities. In these cases, the partition of the unity condition is not fully satisfied:

\begin{equation}
    \sum_{b=1}^{n_b} ~ \frac{m_b}{\rho_b} \kernel \ne 1
    \,.
\end{equation}

The errors in the normalization constraint would introduce a level of uncontrolled errors in the remaining SPH equations. To reduce the normalization errors, the most obvious recipe is to renormalize the kernel itself: 

\begin{equation}
    \rho_a = \sum_b m_b\left(\frac{\kernel}{\sum_{c} \frac{m_c}{\rho^0_c} W_{ac}(h_a)}\right)
    \,,
\end{equation}

\noindent
where  $\rho^0_c = \sum_b m_b W_{cb}(h_c)$ is the standard density. A more clever, albeit more complex, variation of that scheme was developed by \citet{colagrossi03} within the MLS approach, which exactly reproduces the linear variation of a density field (see \citet{gomez10} for a review of these topics). Nevertheless, none of those SPH schemes are totally compatible with the Lagrangian formulation of SPH. The Lagrangian formulation of the equations of movement has the advantage that the grad-h terms can be consistently incorporated to the scheme, so that a complete preservation of mass, linear and angular momentum, energy, and entropy is guaranteed. Most MLS applications belong to the realm of computational fluid dynamics (CFD), which usually works with incompressible or weakly-compressible fluids. Having an almost constant density field for the entire simulation implies that the smoothing length also remains constant and the grad-h corrections are negligible (except at the boundaries, which are usually handled with special techniques). For that reason, the MLS methods are mostly used in CFD simulations; however, in the case of astrophysical scenarios, where density contrasts of orders of magnitude are not rare, the Lagrangian approach with a self-consistent treatment of the grad-h terms is preferable. Nevertheless, due to its properties, we are convinced that the methods presented in this paper can be also of use for CFD simulations.

Other options for the volume elements, $V_a$, may be of interest to address specific problems. The code SPHYNX\footnote{\url{https://astro.physik.unibas.ch/sphynx}} from \citet{cabezon2017}, makes use of the concept of generalized volume elements  \citep{hop13,sai13}. First, a scalar estimator $X_a$ is defined so that the particle volume is: 

\begin{equation}
    V_a=\frac{X_a}{k_a}\,,
    \label{VE_1}
\end{equation}

 \noindent with $k_a=\sum_b X_b W_{ab}(h_a)$. The density of the particle is then calculated as $\rho_a=m_a/V_a$. Current choices for the estimator that can be found in the literature are $X_a=1, m_a,P_a^k$, where P is the pressure and $k\le 1$. There is, however, a particular choice which, according to Paper I, provides a better partition of the unity, namely:

\begin{equation}
    X_a =\left(\frac{m_a}{\rho_a}\right)^p\,.
    \label{estimator}
\end{equation}

\noindent
Setting $p=0$ produces the standard volume element for particles with identical mass, whereas $0 < p\le 1$ gradually improves the kernel normalization when $p\rightarrow 1$.

There are two ways to implement the estimator in Eq.~(\ref{estimator}). The first is to make use of the density calculated in the previous time-step, $\rho_a^{n-1}$, to estimate the volume elements and density at the current iteration $n$:

\begin{equation}
\rho_a^{n}= \frac{m_a \sum_b X_b^{n-1} \kernel}{X_a^{n-1}}\qquad\mathrm {with} \qquad X_a^{n-1} =\left(\frac{m_a}{\rho_a^{n-1}}\right)^p\,.
\label{X1}
\end{equation}

The second is to make use of the density calculated in the standard way in the current time-step, $\rho^0_a=\sum_b m_b \kernel$, to estimate $X_a$ and then  $\rho_a^{n} = m_a/V_a^{n}$ ($V_a^{n}$ calculated with Eq.~\ref{VE_1}). That is:

\begin{equation}
    X_a =\left(\frac{m_a}{\rho^0_a}\right)^p\,.
    \label{X2}
\end{equation}

We refer hereafter to the first method as an {\sl explicit} procedure driven by $X_{1,a}$ (Eq.~\ref{X1}), an to the second as an implicit method, driven by $X_{2,a}$ (Eq.~\ref{X2}).

In the simple, static, toy models discussed below, $X_{1,a}$ convergence is achieved after a few iterations. Once the estimator has converged, there is always an enhancement of the partition of unity, which is almost perfect when the exponent $p\rightarrow 1$. Thanks to its simplicity and good results, $X_{1,a}$ was the estimator chosen in Paper I. It has, however, a couple of drawbacks that are worth noting. Firstly, because of the explicit nature of Eq.~(\ref{X1}), a complete Lagrangian consistency is never achieved. Secondly, taking $p=1$ makes it too sensitive to particle noise and not recommended. Therefore, $p=0.7-0.8$ are the recommended values, which slightly degrades the performance of $X_{1,a}$ and introduces an undesirable free parameter (the particular value of the exponent $p$). These issues can be overcome with the second procedure to implement $X_a$ which, in the end, was the one chosen in this work. Therefore,  $X_a$ consistently refers  to $X_{2,a}$ hereafter.
 
In the second method, the calculation of the VEs makes use of the value of $\rho^0_a$ calculated in the current integration step. As we go on to see in the static toy models (Sect.~\ref{sec:static}), the explicit option $X_{1,a}$ leads to a better partition of the unity and interpolations than $X_{2,a}$ for identical values of the $p$~exponent, but the former is less numerically robust and not fully compatible with the Lagrangian formulation of the SPH equations. On the contrary, the estimator $X_{2,a}$ allows us to build a Lagrangian-consistent scheme (Appendix~\ref{sec:Appendix}), which incorporates the grad-h terms -- provided that the exponent $p$ in Eq.~(\ref{X2}) is chosen equal to one ($p=1$). This allows us to eliminate a parameter.

In the following sections, we show that reducing the error in the kernel normalization ($E_1$, hereafter) of particle $a$, as in:
\begin{equation}
E_1=\left[\sum_b V_b \kernel-1\right]\,,
\label{errorE1}
\end{equation}

\noindent
usually improves the requirement:

\begin{equation}
E_2\cdot h_a = \left|\left<{\bf \Delta r}\right>\right|_a = \left|\sum_b V_b~({\bf r}_b-{\bf r}_a)~\kernel\right|\simeq 0\,,
\label{errorE2}
\end{equation}

\noindent
where $E_2$ is the normalized error module ${\bf \left|\left<\Delta r \right>\right|}_a/h_a$ of particle $a$. The error reduction is the consequence of using the estimators $ X_{1,2}$ to evaluate the volume elements. Any reduction in both errors ($E_1, E_2$) will potentially improve the dynamic evolution of the simulated physical system.

\section{Estimating the errors \texorpdfstring{$E_1$}{E1} and \texorpdfstring{$E_2$}{E2} for different particle distributions}

\label{sec:static}

A good control of errors $E_1$ and $E_2$ is of utmost importance to the SPH technique. This is because the quality of both the interpolated function $\left<f(\bf{r})\right>$,

\begin{equation}
    \left<f({\bf r})\right>  \simeq f({\bf r})\sum_b V_b \kernel + {\bf\nabla f}\cdot \sum_b V_b ({\bf r}_b-{\bf r}_a)\kernel\,,
    \label{intfunc}
\end{equation}
\noindent
and its gradient $\left< {\bf \nabla}f({\bf r})\right>$ (Eq.~\ref{matrix}), are very sensitive to these errors \citep{ros15}. Nevertheless, both errors are correlated, as it can be inferred from the following argumentation in one dimension. We first take the kernel normalization condition as a function of the spatial coordinate,

\begin{equation}
    G(x_a)=\sum_b V_b \kernel\,.
\end{equation}

Using the Gaussian kernel $\kernel= \frac{C}{h_a}~\exp[-(\frac{x_b-x_a}{h_a})^2]$, the standard SPH derivative of $G(x)$ is expressed as:

\begin{equation}
 \left(\frac{dG}{dx}\right)_a=-\frac{2}{h_a^2}\sum_b V_b  (x_b-x_a) \kernel\,,
    \label{derivG1}
\end{equation}

\noindent
thus,

\begin{equation}
    \sum_b V_b  (x_a-x_b) W_{ab} = -\frac{h_a^2}{2} \left(\frac{dG}{dx}\right)_a\,.
    \label{derivG2}
\end{equation}

We note that the LHS of Eq.~(\ref{derivG2}) is in fact $E_2$, suggesting that having a good partition of the unity ($G\simeq 1$, i.e., $E_1\to 0$) makes $dG/dx \simeq 0$ and, as a consequence, the $E_2$ error is suppressed. An independent proof of the link  between $E_1$ and $E_2$ obtained with an exponential kernel \citep{fulk96} is given in the Appendix A of Paper I.

It should be recognized, however, that the proof above is only indicative because, unlike the interpolators widely used in practical calculations, the Gaussian and the exponential kernels are functions without compact support. Moreover, it could happen that even when $G(x)$ is close to one, it  shows fluctuations around the particles and its derivative may significantly differ from zero.

In this regard, additional insights about the impact of the VE choice in the errors $E_1$ and $E_2$ can be obtained by studying a handful of static particle distributions and using kernels with compact-support. These errors are expected to be large close to discontinuities, which in SPH are usually spread over a few times the smoothing-length distance ($h$). We chose three representative discontinuities that often appear in practical calculations: a Gaussian (model A), an inverted-Gaussian (model B), and a wall (model C). These are given by the following mathematical expressions:

\begin{equation}
\rho(x)=\rho_0+\Delta\rho~e^{-(\frac{x-x_0}{\delta})^2}, \qquad (A)
\label{mountain}
\end{equation}

\begin{equation}
\rho(x)=\rho_0-\Delta\rho~e^{-(\frac{x-x_0}{\delta})^2},\qquad (B)
\label{valley}
\end{equation}

\begin{equation}
\rho(x)=\rho_0+\Delta\rho~\frac{e^{(\frac{x-x_0}{\delta})}-e^{-(\frac{x-x_0}{\delta}})}{e^{(\frac{x-x_0}{\delta})}+e^{-(\frac{x-x_0}{\delta}})}, \qquad (C)
\label{wall}
\end{equation}

\noindent
where the values of the parameters $\rho_0$, $\Delta\rho$, and $\delta$ are specified in Table~\ref{table1}.

\begin{figure*}

\includegraphics[width=\textwidth]{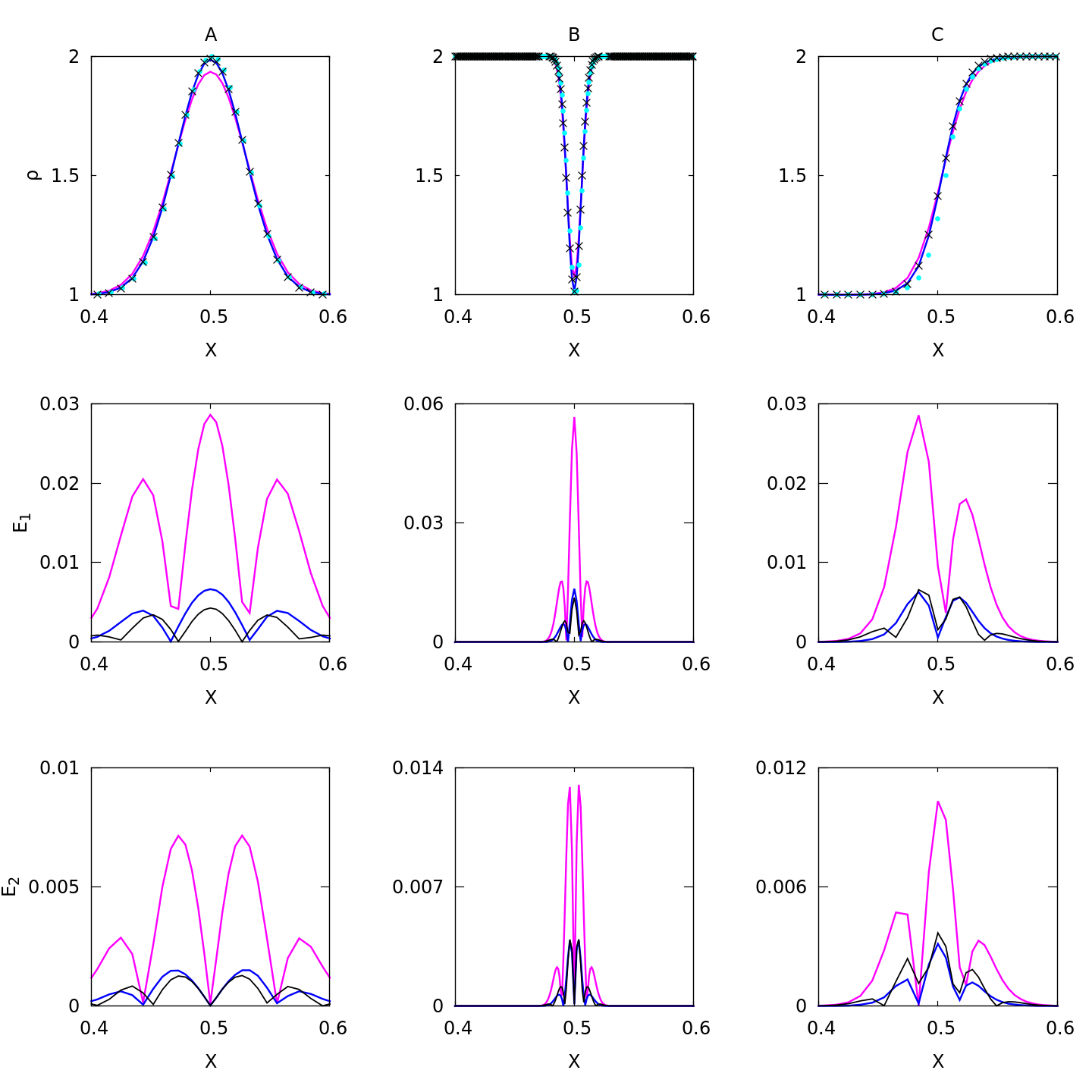}
    \caption{Results of the SPH evaluation of the test profiles A, B, and C using different VEs. Top row: Density profiles of models A (Gaussian), B (Inverted-Gaussian), and C (Wall), of Table~\ref{table1}, calculated with $X_1$ using $p=0.0$ (magenta) and $p=0.8$ (blue) in Eq.~(\ref{estimator}). Crosses ($\times$) in black are for $X_2$ (Eq.~\ref{X2}, with $p=1$) and points in light blue are the analytic values. Central row: $E_1$ (partition of unity) error for the same three profiles following the same color schemes, with the black lines for $X_2$ with $p=1$. Bottom row: Same as central row but for error $E_2={\bf\vert\left<\Delta x\right>\vert}/h$.}
    \label{fig:errorsfunc}
\end{figure*}

We arranged these profiles into a 1D distribution of $N=100$ particles  with equal mass and distributed according to the density profile, with reflective boundary conditions. We used an interpolating $sinc$-kernel with exponent $n=5$ \citep{cabezon2008}, which has a shape similar to the $M_6$ spline. The different profiles of errors $E_1$ and $E_2$ at each point, for the estimators $X_1$ and $X_2$, are depicted in Fig.~\ref{fig:errorsfunc}. As we can see, $X_1$ leads to a clear improvement in the kernel normalization condition as $p$ increases. For $p\simeq 1$ the $E_1$ error becomes negligible, hence it is not shown. Interestingly, the estimator $X_2$ with $p=1$ (black crosses) leads to similar $E_1$ and $E_2$ errors as $X_1$ with $p= 0.8$ in all profiles. The normalized error $E_2 = {\bf\vert\left<\Delta x\right>\vert}/h$ follows a similar trend, suggesting that having a good partition of unity is not only beneficial to approaching an estimaton of the density, but also to calculate the gradient of any magnitude of interest with ISPH.

\begin{table}[ht!]
        \centering
        \caption{Value of the different parameters in profiles A, B, and C that mimic different types of sharp density gradients. Profile D is the same as C but with different parameters.}
        \begin{tabular}{ccccc} 
                \hline
                Profile & $\rho_0$ & $\Delta\rho$ & $\delta$ & h\\
                \hline
                \hline
                A & 1.0 & 1.0 & 0.040&0.0230\\
                B & 2.0 & 1.0 & 0.008&0.0051\\
                C & 1.5 & 0.5 & 0.020&0.0230\\
                D & 10 & 11 & 0.040&0.0230\\
                \hline
        \end{tabular}
        \label{table1}
\end{table}

\subsection{Impact of VE in estimating gradients}
\label{sec-vegrad}
We can use our simple sharp profiles above to gain insight into the relationship between $E_1$,  $E_2$ in Eqs.~(\ref{errorE1}, \ref{errorE2}), as well as the accuracy of the first derivative. To do so, we assume that the density of the test particle distribution follows profile A, Eq.~(\ref{mountain}), so that it totally determines the VEs through Eq.~(\ref{estimator}). Let us also assume  that we wish to obtain the SPH derivative of a generic wall-like function $f$ given by profile $D$ in Table~\ref{table1}, Eq.~(\ref{wall})\footnote{This test would mimic, for example, a thermal wave passing through a star.}. Such a derivative, $\frac{df}{dx}$, is sensitive to the choice of the estimator $X$ to compute the VEs. We can thus compare the analytic and the numerical value of $\frac{df}{dx}$ and carry out the $L_1$ analysis of the results  with Eq.~(\ref{L1formula}). 

\begin{figure}
\centering
\includegraphics[width=\columnwidth]{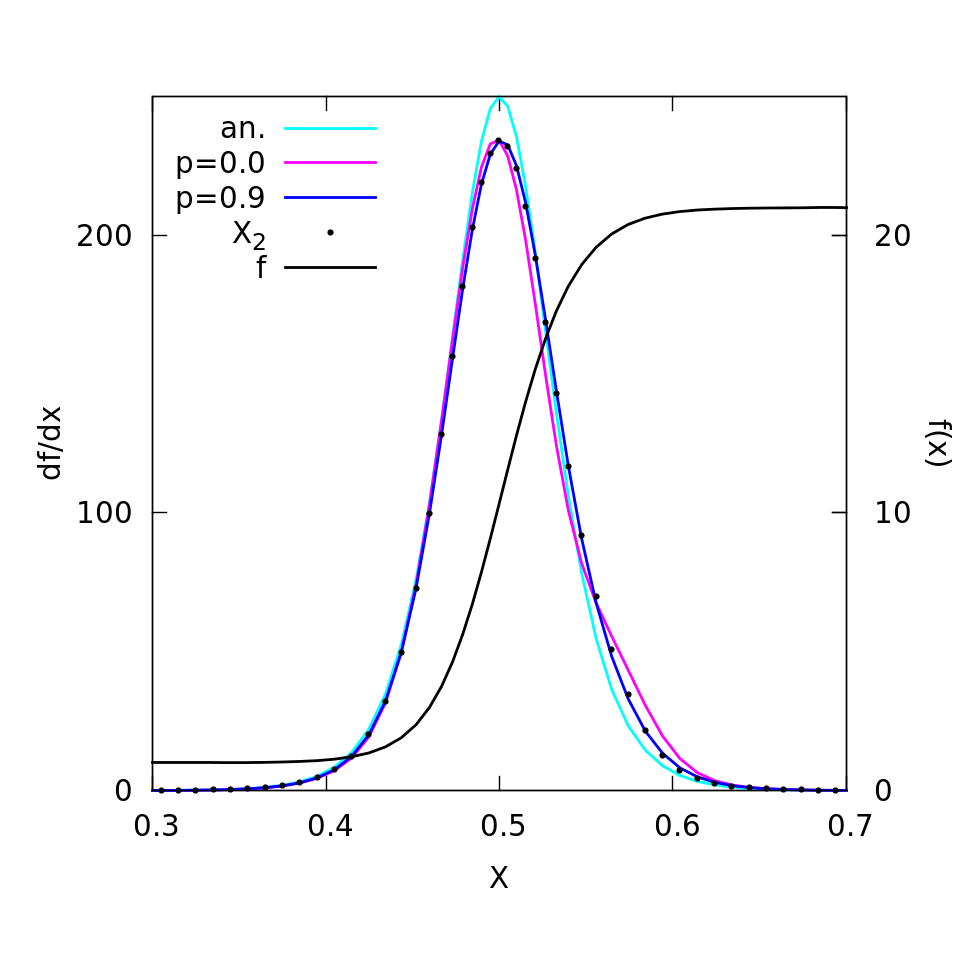}
    \caption{Wall-like function f(x) (black line) and its analytical gradient $df/dx$ (light-blue solid line). The solid lines in magenta and blue were obtained with $X_1$ and $p=0.0$, 0.9 respectively, while the black points are for $X_2$ with $p=1$. }
    \label{fig:fittingVE}
\end{figure}

Figure~\ref{fig:fittingVE} depicts the wall-like function $f(x)$ (black solid line) and its analytical gradient of $df/dx$ (light-blue line). Although the value around the maximum of $\frac{df}{dx}$ is similar for $p=0.0$ (magenta solid line) and $p=0.9$ (blue solid line), the last option fits better the derivative around the coordinate $x=0.6$, which is in turn similar to that obtained using the estimator $X_2$ with $p=1$ (black dots).

\begin{figure*}
\centering
\includegraphics[width=\textwidth]{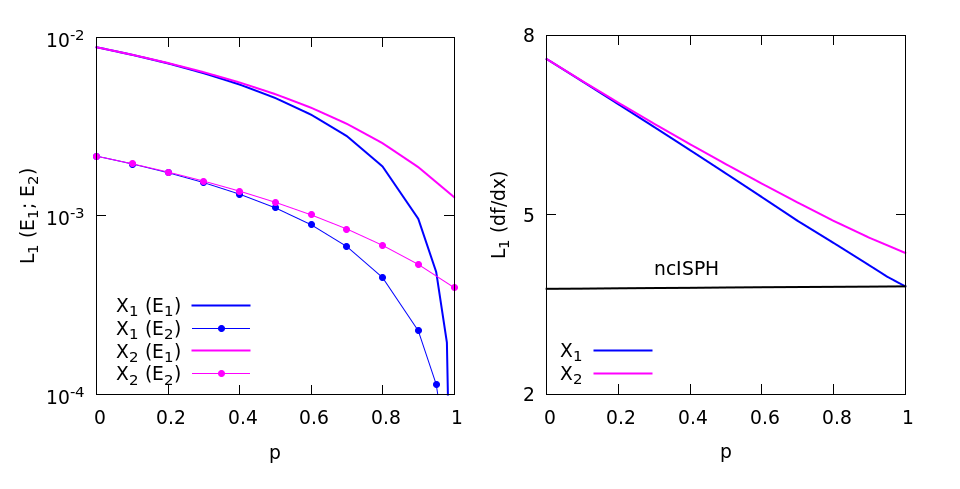}
    \caption{$L_1$ calculation of the errors in the numerical experiments shown in Fig.~\ref{fig:fittingVE}. Left: Averaged value of $L_1$ for the partition of the unity (solid lines) and $\left<\Delta x\right>/h$ (dotted lines). Blue lines are for the estimator $X_1$ and magenta lines for $X_2$. Right: averaged value of $L_1$, obtained with Eq.~(\ref{L1formula}), for the derivative of the wall function calculated with $X_1$ (solid blue line) and $X_2$ (solid magenta line). The black line is the calculation with the non-conservative ISPH, which is independent of the particular value of the exponent $p$. We note that when $p\simeq 1$ both schemes (ISPH and ncISPH) converge, albeit faster for estimator $X_1$ than for $X_2$. }
    \label{fig:convergence}
\end{figure*}

The left panel in Fig.~\ref{fig:convergence} shows the averaged $L_1$ value for $E_1$ (Eq.~\ref{errorE1}) and $E_2$~(Eq.~\ref{errorE2}): 

\begin{equation}
    L_1 (E_{1,2})= \left<\vert E_{1,2}\vert\right >,
\label{L1E_1E_2}
\end{equation}

\noindent calculated in the interval $0.3\le x\le 0.7$ for the test presented above. The $L_1$ values for the partition of the unity and $\left<\Delta x\right>/h$ decreases as the exponent p increases, as expected. Nevertheless, the quantitative details again depend on the type of estimator, $X_1$ or $X_2$. As the figure shows, there is a factor ten reduction of the $L_1$ errors for $X_1$ and $p\simeq 0.9$. The errors become negligible when $p\simeq 1$. Although the option $X_2$ shows a lower convergence rate at larger p values, it still provides a significant reduction of  $L_1$ , decreasing the errors in almost a factor ten when $p= 1$. 

The panel on the right in Fig.~\ref{fig:convergence} shows the averaged $L_1$ error of $\frac{df}{dx}$ normalized to $\left<(\frac{df}{dx})_{analytic}\right>$, 

\begin{equation}
    L_1 =\frac{1}{N \left< (\frac{df}{dx})_{analytic})\right>}\sum_{b=1,N} \left\vert\left(\left(\frac{df}{dx}\right)_{analytic}-\left(\frac{df}{dx}\right)_{sph}\right)_b\right\vert
    \,,
    \label{L1formula}
\end{equation}

\noindent
and how both schemes (ISPH and ncISPH) converge in light of the flavor used for the estimator $X$ ($X_1$ solid blue line and $X_2$ solid magenta line).
We note that because the ncISPH scheme (Eq.~\ref{approxI})  makes exact linear interpolations, the results (black line) are  not sensitive on the adopted value of $p$. On the other hand, the results of the conservative scheme, ISPH (blue and magenta lines), show a clear dependence on the choice of the volume element, as expected. The profile of $L_1 (p)$ decreases linearly, achieving the same accuracy as the ncISPH scheme when $p > 0.9$ for $X_1$ or close to it for $X_2$. 

Even though the test cases presented in this section were calculated in 1D with ordered particle settings, the results unambiguously support the idea that: a better partition of unity improves both the estimation of density and the calculation of gradients in ISPH. Hereafter, we focus on the volume elements calculated with the estimator $X_2$ (Eq.~\ref{X2}) with $p=1$, because, unlike $X_1$, it  fits in with the Lagrangian formulation of the SPH equations perfectly and also allows us to eliminate the parameter, $p$. Additionally, the grad-h terms can be easily incorporated to the scheme with this choice of VEs. The resulting ISPH equations are described in Sect.~\ref{sec:isph_equations}.    
 
\subsection{The quest for fully implicit VEs}

In principle, the best procedure for calculating the volume elements would be to directly obtain them from the inversion of the kernel normalization matrix. Unlike the indirect methods described in the previous section, a fully implicit implementation of the VEs has the advantage of always leading to the perfect partition of the unity. Previous attempts to build implicit SPH schemes \citep{Knapp2000, escartin2016}, have made use of advanced techniques to efficiently invert large sparse matrices, as, for example, the PARDISO library\footnote{The widely used Intel MKL PARDISO library function is based on a legacy version of this project: \url{https://www.pardiso-project.org/}}. These types of libraries could also be used to implicitly find the volume elements $V_b$ by solving large linear systems with $N\times N$ equations and unknowns:

\begin{equation}
\sum_{b=1}^{n_b}~V_b~ W_{ab}(h_a)=1\,.
\label{VEdirect}
\end{equation}

\begin{figure}
\centering
\includegraphics[width=\columnwidth]{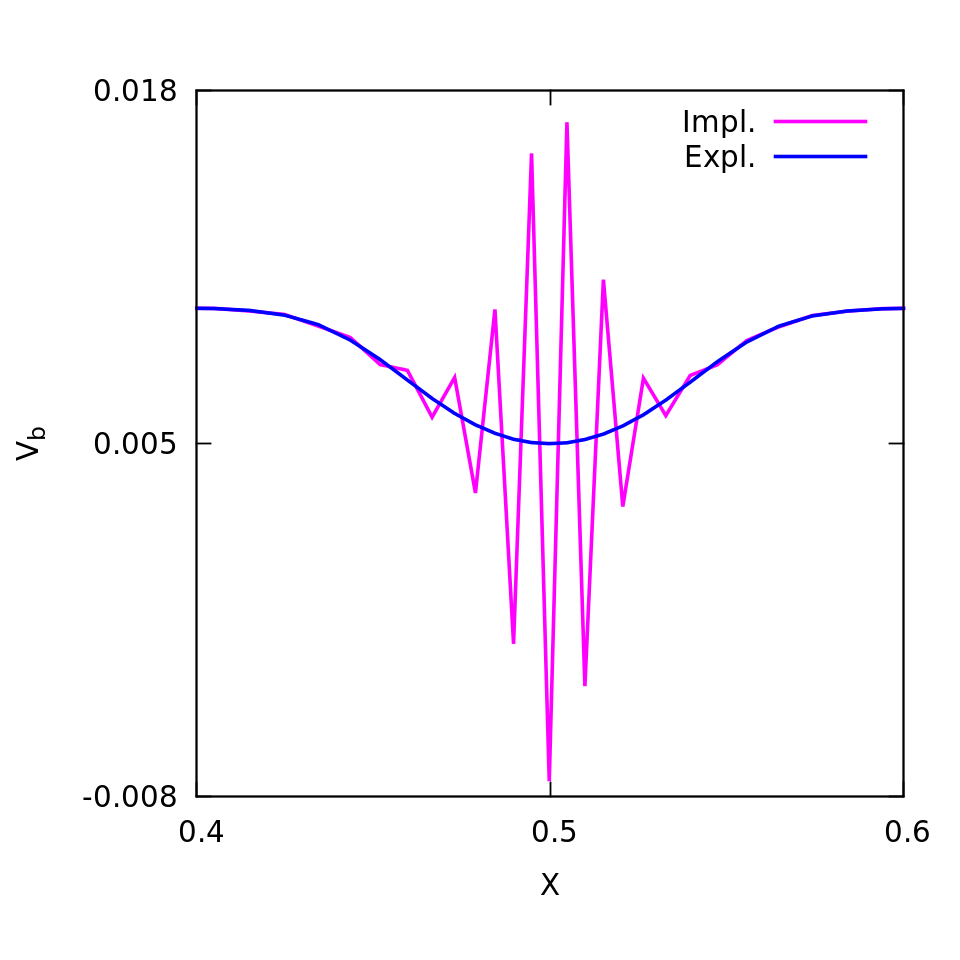}
    \caption{Implicit versus explicit estimation of the volume elements, $V_b$~, in the numerical experiments with the Gaussian curve (Eq.~\ref{mountain}).} 
    \label{fig:implicitexpplicit}
\end{figure}

Unfortunately, finding the VEs with such a direct approach generally leads to non-physical results. We calculated, in a fully implicit manner, the VEs of the Gaussian function (Eq.~\ref{mountain}) by solving the linear system above and found that the volume elements strongly oscillate around the explicit solution given by Eq.~(\ref{X1}) with $p=1$ (see Fig.~\ref{fig:implicitexpplicit}). Furthermore, in many points, the volume elements become negative, which is also non-physical. Therefore, we discarded the fully implicit route to finding the VEs in this work.

\section{ISPH equations}
\label{sec:isph_equations0}
Here, we summarize the set of ISPH equations used to compute the 3D tests on the following section. We provide two sets of momentum and energy equations, one corresponding to a "standard" of generally used VEs (i.e., $X_a=m_a$) and another one to our suggested VEs, which are the equations used to perform the comparisons in the tests of Sect.~\ref{sec:dynamic}. We also provide the details of the procedure used to build these equations in Appendix~\ref{sec:Appendix}. 

\subsection{Description of the ISPH equations}
\label{sec:isph_equations}

 The density equation is expressed as:  
\begin{equation}
    \rho_a = \frac{m_a}{V_a}\,,
    \label{density_1}
\end{equation}
with $V_a$~defined by Eq.~({\ref{VE_1})}.

    The momentum equation is:
\begin{equation}
  \ddot{x}_{i,a}=
  \begin{cases}
  \mathrm{For}~X_{a,b}=m_{a,b}:\\
  -\frac{X_a}{m_a}\sum_b m_b\left[\frac{X_b P_a}{\Omega_a k_a^{2-\sigma} k_b^{\sigma}} \mathcal A_{i,ab}(h_a) + \frac{X_b P_b}{\Omega_b k_b^{2-\sigma}k_a^{\sigma}}\mathcal A_{i,ab}(h_b)\right];\\  \\
  
  \mathrm{For}~X_{a,b}=m_{a,b}/\rho_{a,b}^0:\\
  - \sum_b m_b\left[\frac{X_a^{2-\sigma}X_b^{\sigma} P_a}{\Omega_a m_a^2~k_a} \mathcal A_{i,ab}(h_a) + \frac{X_b^{2-\sigma}X_a^{\sigma} P_b}{\Omega_b m_b^2~k_b}\mathcal A_{i,ab}(h_b)\right];
  \end{cases}
  \,,
  \label{mom_1}
\end{equation}

\noindent with $\rho^0_a=\sum_b m_b \kernel$, and $\Omega_{a,b}$ is given in Appendix~\ref{sec:Appendix}.  We introduced a parameter, $0\leq\sigma\leq 1$,~which allows us to choose between the pure Lagrangian scheme developed in the Appendix~\ref{sec:Appendix} ($\sigma=0$) and a progressive deviation of it, which suppresses the tensile instability when $\sigma\to 1$ (see Section  \ref{subsec:sigma} for more details on $\sigma$).  Moreover,  

\begin{equation}
    \mathcal A_{i,ab}(h_{a,b})=\sum^d_{j=1} c_{ij,a}(h_a)(x_{j,b}-x_{j,a})W_{ab}(h_{a,b})
    \,,
\end{equation} 
  
\noindent with $c_{ij,a}$ as the coefficients of the inverse matrix in the IA given by Eq.~(\ref{matrix}), and $d$ as the number of dimensions. Any expression of standard SPH can indeed made compatible with the IA by taking the kernel derivative as \citep{cabezon2012}:

\begin{equation}
    \frac{\partial W_{ab}(h_a)}{\partial x_{i,a}}= \mathcal A_{i,ab}(h_a);\quad i=1,d
    \,.
    \label{stdiad}
\end{equation}

    The energy equation is:

\begin{equation}
  \dot{u}_a=
  \begin{cases}
  \mathrm{For}~X_{a,b}=m_{a,b}:\\
  \frac{X_a P_a}{m_a\Omega_a k_a^{2-\sigma}}\sum_b\sum^d_{i=1}  \frac{X_b}{k_b^{\sigma}}\left[(v_{i,a}-v_{i,b}) \mathcal A_{i,ab}(h_a)\right];\\ \\
  \mathrm{For}~X_{a,b}=m_{a,b}/\rho_{a,b}^0:\\
  \frac{X_a^{2-\sigma} P_a}{m_a^2\Omega_a k_a}\sum_b\sum^d_{i=1} m_b X_b^{\sigma}\left[(v_{i,a}-v_{i,b}) \mathcal A_{i,ab}(h_a)\right];
  \end{cases}
  \,.
  \label{energy_1}
\end{equation}

The necessary terms of artificial viscosity (AV) are added to the right of the equations above as in Paper I, but explicitly including a quadratic term in the signal velocity:

\begin{equation}
    v_{ab}^{sig} = \bar\alpha_{ab}\bar c_{s,ab}-\beta w_{ab}
    \,,
\end{equation}

\noindent
where $w_{ab}={\bf v}_{ab}\cdot {\bf \hat r}_{ab}$ \citep{mon97,price18}. The parameter $\beta$ is kept constant with a default value $\beta=2$. The AV coefficient $\alpha$ is controlled with the switches scheme described in \citet{rea10} so that $\alpha\simeq 1$ in strong shocks but it decays to a minimum value of $\alpha\simeq 0.05$ away from them. Finally, $\bar\alpha_{ab}$ and $\bar c_{s,ab}$ are the average $\alpha$ and speed of sound between neighboring particles. According to \citet{mon97}, the AV contribution to the energy equation should include a heat-conduction term which smooths the pressure in wall-shock conditions. The precise form of such heat-conduction term is:

\begin{equation}
    \left(\frac{du_a}{dt}\right)_{cond}^{AV}=\sum_b\sum_{i=1}^d m_b\alpha_u \frac{v_{ab,cond}^{sig} (u_a-u_b)}{\bar\rho_{ab}}  \frac{r_{i,ab}}{\vert r_{ab}\vert} \bar{\mathcal A}_{i,ab}(h_a, h_b)
    \,,
    \label{conduction}
\end{equation}

\noindent
with $\bar{\mathcal A} (h_a,h_b)=0.5 (\mathcal A (h_a)+\mathcal A(h_b))$, which in the SPHYNX code is implemented as: 

\begin{equation}
\begin{split}
    \left(\frac{du_a}{dt}\right)_{cond}^{AV}=&  
    \sum_b\sum_{i=1}^d \frac{1}{2}~\alpha_u~v_{ab,cond}^{sig} (u_a-u_b)\\
    &\left\{V_a~\frac{m_b}{m_a} \frac{r_{i,ab}}{\vert r_{ab}\vert} \mathcal A_{i,ab}(h_a)+V_b~ \frac{r_{i,ab}}{\vert r_{ab}\vert} \mathcal A_{i,ab}(h_b)\right\}
    \end{split}
    \,.
    \label{conductionSphynx}
\end{equation}

The signal velocity $v^{sig}_{cond}$ used in our tests is \citep{price08}: 

\begin{equation}
v_{ab,cond}^{sig}= \sqrt{\frac{\vert P_a-P_b\vert}{\bar\rho_{ab}}}
\,.
\end{equation}

The results of the tests below suggest that adding a small amount of conductive heat by Eq.~(\ref{conductionSphynx}) is beneficial because it contributes to reduce the tensile instability and to smooth the numerical noise. Nonetheless, the value of the constant $\alpha_u$ should not be high, otherwise the density-peak in strong shocks may be under estimated (see Section \ref{subsec:sedov}).  We chose  $\alpha_u=0.1$, which is low and in agreement with the choice by other authors, such as ~\citet{tricco19}. The equation of state (EOS) is that of an ideal gas, with $\gamma=5/3$.

\subsection{Obtaining the value of \texorpdfstring{$\sigma$}{[sigma]}}
\label{subsec:sigma}

When using the SPH equations deduced from the Euler-Lagrange formulation a drawback appears wherever $\nabla\rho$ becomes large as in the case of  contact discontinuities, for instance. In such cases, the incorrect estimation of gradients may lead to numerical artifacts, with the tensile instability standing as one of the most harmful and most common. Several solutions have been postulated to cope with this problem, all of them requiring some departure from the exact Lagrangian formulation. For example, \citet{rit01} proposed to estimate the density by averaging over the internal energies so that the ensuing density field is smooth. A similar approach was described in \citet{sai13}, who suggested that the volume elements be redefined so that they depend on the pressure rather than density. Another solution was proposed by \citet{rea10}, where a typical element within the summations of the momentum and energy equations is changed from the standard $[P_a/\rho_a^2]$ to $[P_a/(\rho_a\rho_b)]$. While it is not totally Lagrangian compatible, such a simple change is mathematically consistent with the standard derivation of the SPH equations \citep{mon92,price04}\footnote{The possibility of working with $P_a/(\rho_a^{2-\sigma} \rho_b^{\sigma})$ (with some preference for $\sigma=0$) was already suggested by J. Monaghan (1992) in his popular review, but $\sigma$ was thought to be a constant real number.}  and can totally suppresses the tensile instability. All of these SPH variants  have a feature in common, namely, that the main magnitudes in the movement equation are somehow "crossed" for particles $a$ and $b$ [f.e. $u_b/\rho_a$ in \citet{rit01} or $P_a/(\rho_a\rho_b)$ in \citet{rea10}]. In \citet{cabezon2017}, the choice of $X_{1,a}$  (Eq. \ref{X1}), combined to the SPH equations defined in \citet{ros15b}, naturally led to an almost "crossed" scheme\footnote{Specifically, it can be shown that the scheme in Paper I involve terms such as $P_a/(\rho_a^{2-p}\rho_b^p)$, with $p=0.7$~being the exponent of the explicit estimator (Eq. \ref{X1}) considered in that work.}. In spite of not being totally Lagrangian-compatible and with the grad-h terms only approximated, this code behaved well and was able to pass a suite of verifying tests.

Here, we propose a similar procedure as in \citet{rea10}, but allowing for a self-adaptive crossing of indexes in the original Lagrangian equations derived in the Appendix~\ref{sec:Appendix}. The resulting expressions, Eqs.~(\ref{mom_1}) and (\ref{energy_1}), are steered by a parameter $\sigma$ so that $0\leq \sigma\leq 1$. The value $\sigma=0$ leads to the fully Lagrangian SPH equations whereas $\sigma=1$ gives a fully crossed expression. We make use of a ramp function ($R$) to automatically decide the instantaneous value of $\sigma$ as a function of the density contrast between any pair of particles.

\begin{equation}
\sigma_{ab}=
\begin{cases}
0; &  At_{ab}\leq At_{min}\\
R (At_{ab}-At_{min}); & At_{min}\leq At_{ab}\le At_{max}\\
1; & At_{ab}\geq At_{max}\,
\end{cases}
\,,
\label{ramp}
\end{equation}

\noindent
with $R=\frac{1}{At_{max}-At_{min}}$ and $At_{ab}=\left\vert\frac{\rho_a-\rho_b}{\rho_a+\rho_b}\right\vert$ as the Atwood number. Equation  (\ref{ramp}) leads to $\sigma\simeq 1$ only in those regions hosting large density gradients, while the fully Lagrangian scheme is preserved wherever $\sigma =0$, which is taken by the majority of the particles of the system. We empirically found that $At_{min}=0.1, At_{max}=0.2$ produce satisfactory results in all the numerical experiments described in this work. 

\section{Tests}
\label{sec:dynamic}
In this section, we analyze the results of four well-known tests that require the solution of the full system of Euler equations in three dimensions. The first test deals with the hydrostatic equilibrium of a two-phase fluid. It aims to analyze the abilities of our Lagrangian scheme to handle sharp density discontinuities. Then we applied this scheme to the study of the growth of the Kelvin-Helmholtz (KH) instability, to  the interaction of a supersonic wind with an isothermal, high-density spherical cloud (the wind-cloud test), as well as to simulate the evolution of a point-like explosion (the Sedov test). The simulations were carried out using the ISPH code SPHYNX and the outcomes have been compared to well-known solutions. Tests were run with varying parameters to  disentangle the effects of IA, VE, heat transport in the AV, and $\sigma$. Special emphasis is placed on a comparison among models with different choices of $\sigma$ and volume elements, namely, the standard choice\footnote{Actually, any constant magnitude is a suitable choice as $X_a$,~but taking the mass of the particle allows for the fine-tuning of the density (if needed)  by slightly modifying the mass of the particles. In nearly isobaric systems, with shallow pressure gradients, the choice $X_a=P_a^k,~  k\le 1$ could also be appropriate.} of $X_a=m_a$ and the improved VEs with $X_a=m_a/\rho^0_a$.

 We find that the conservation properties are excellent and that linear momentum is preserved to almost machine precision in all tests. The angular momentum, with respect to the center of mass, deviates less than $10^{-6}$~from the initial value. The relative deviation, $\vert\Delta E\vert/E_0$ in the total energy was less than $10^{-5}$, except in the isobaric two-fluid test (Sect.~\ref{subsec:isobaric}), where it was $\simeq 10^{-4}$~at late times.

\subsection{Isobaric two-fluid test}
\label{subsec:isobaric}

The simulation of the hydrostatic evolution of a two-phase system with very different densities and internal energies is far from being trivial with SPH codes. We considered a system of two fluids separated  by a contact discontinuity but in pressure equilibrium:

\begin{equation}
\rho=
\begin{cases}
4 & 0.25\le x, y, z\le 0.75\,,\\
1 & \mathrm{otherwise}\,.
\end{cases}
\label{twofluid}
\end{equation}

The system is isobaric with $P=2.5$ and we use $N=110^3$ equal-mass particles  spread in two nested body-centred cubic (bcc) lattices. The EOS is $P=(\gamma-1) \rho u$ with $\gamma=5/3$, the number of neighbors is set to $n_b=100,$ and the $sinc~(n=5)$ kernel  \citep{cabezon2008} is used to interpolate. If the density around the contact discontinuity is not adequately smoothed, the two phase system  evolves non-physically when the full Lagrangian scheme (Eqs.~\ref{mom_1} and \ref{energy_1}, with $\sigma=0$) is applied.  The reason is that the error in $\nabla\rho$ becomes too large at the contact discontinuity, inducing the tensile instability. This is just the kind of simulation where the use of the magnitude $\sigma$ (see Section \ref{subsec:sigma}), defined by Eq.~(\ref{ramp}), becomes especially helpful.

\begin{figure}
\includegraphics[angle=0,width=\columnwidth]{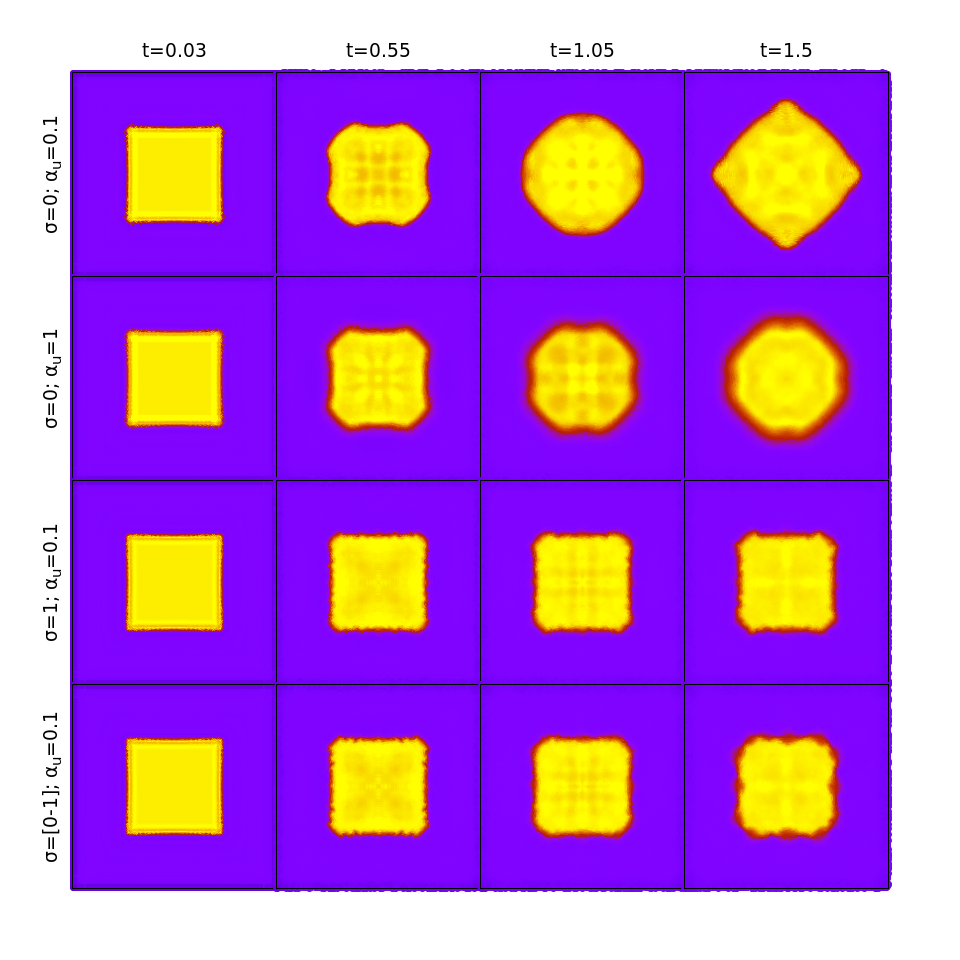}
    \caption{Slices  around $Z=0$ showing the density colormap of models $H_5, H_6, H_7, H_8$\tablefootnote{ Very similar results for H1-H3 models, calculated with standard VEs can be found in plots given in previous works, \cite[their Fig.~4]{cabezon2017}}. in Table \ref{squaretesttable} (rows) at different times (columns), with the sound-crossing time being $\tau_{sc}=0.9$.}
    \label{fig:isobaric_cube}
\end{figure}

\begin{figure}
\centering
\includegraphics[width=\columnwidth]{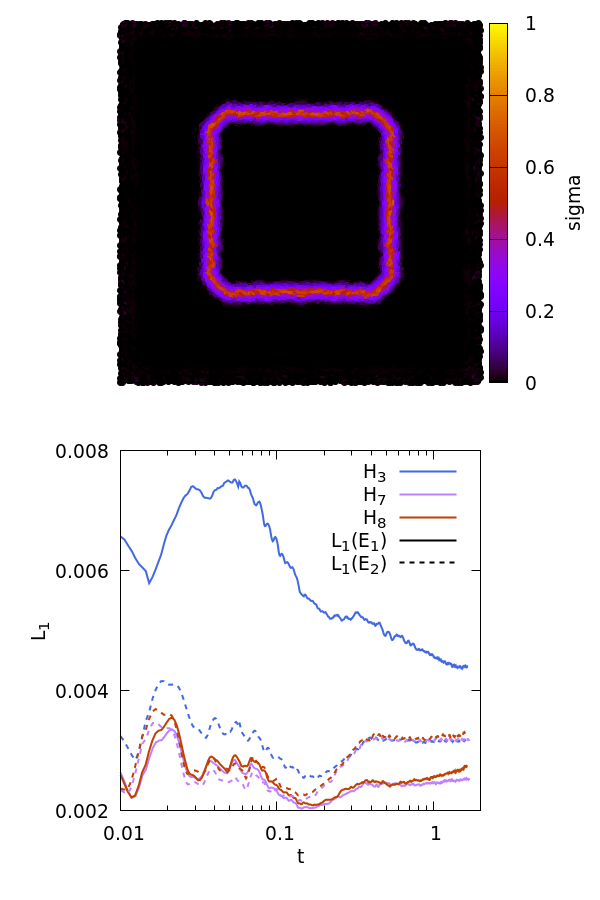}
    \caption{
    Isobaric two-fluid numerical experiment. Upper panel:   Slice around Z=0 showing  the averaged $\sigma$-parameter (Eq.~\ref{ramp}), in model $H_8$ at $t=0.55$. Appreciable values of $\sigma$ are only attained at the fluid inter-phase. Lower panel: Time evolution of averaged $L_1(E_1)$ (solid lines), $L_1(E_2)$ (dashed lines) for models $H_3$, $H_7$, and $H_8$ in Table~(\ref{squaretesttable}).}
    \label{fig:isobaric_sigma}
\end{figure}

\begin{table}
        \centering
        \caption{$L_1$ values for errors $E_1$ and $E_2$ at $t=1.5$ in the hydrostatic square test. The significance of the columns: $\sigma$ is the magnitude defined in Eq.~(\ref{mom_1}), $\alpha_u$ controls the amount of heat transport in the AV (Eq.~\ref{conduction}), $X$ is the estimator connected to the VE choice and $L_1(E_1)$, $L_1(E_2)$, $L_1(d)$ are the averaged $L_1$ values of the partition of unity, the $\left<\Delta r\right>/h$ condition, and the deviations of the SPH particles from their initial position, respectively. }
        
        \begin{tabular}{ccccccc} 
                \hline
         \multirow{2}{*}{Model} & \multirow{2}{*}{$\sigma$} & \multirow{2}{*}{$\alpha_u$} & \multirow{2}{*}{$X$} & $L_1 (E_1)$ & $L_1 (E_2)$ & $L_1(d)$ \\
         &&&&$[\times10^{-3}]$ & $[\times10^{-3}]$ & $[\times10^{-2}]$ \\
                \hline
                \hline
                $H_1$ & $0.0$   & $0.1$ & $m$        & $3.0$ & $2.5$ & $3.2$ \\
                $H_2$ & $0.0$   & $1.0$ & $m$        & $3.0$ & $2.6$ & $3.1$ \\
                $H_3$ & $1.0$   & $0.1$ & $m$        & $4.4$ & $3.2$ & $1.6$ \\
                $H_4$ & $[0-1]$ & $0.1$ & $m$        & $5.0$ & $3.5$ & $2.0$ \\
                $H_5$ & $0.0$   & $0.1$ & $m/\rho^0$ & $2.1$ & $2.7$ & $3.2$ \\
                $H_6$ & $0.0$   & $1.0$ & $m/\rho^0$ & $2.5$ & $2.9$ & $3.0$ \\
                $H_7$ & $1.0$   & $0.1$ & $m/\rho^0$ & $2.5$ & $3.2$ & $1.5$ \\
                $H_8$ & $[0-1]$ & $0.1$ & $m/\rho^0$ & $2.7$ & $3.3$ & $2.0$ \\
                \hline
        \end{tabular}
        \label{squaretesttable}
\end{table}

The results of applying Eqs.~(\ref{mom_1}) and (\ref{energy_1}) to the hydrostatic square test are summarized in Table~\ref{squaretesttable} and Fig.~\ref{fig:isobaric_cube}.  Only models with $X_a=m_a/\rho^0_a$ are shown in Fig.~\ref{fig:isobaric_cube}, and calculations with $X_a=m_a$ give similar results, albeit with higher L1 values. The density color map slices depicted in Fig.~\ref{fig:isobaric_cube} show that the behavior of the square is primarily controlled by the value of $\sigma$. In the case of the full Lagrangian formulation (models $H_5, H_6$), calculated with $\sigma=0$, the system completely looses its shape  in half of the sound-crossing time, $\tau_{sc}\simeq 0.9$ s. Nevertheless, increasing the conductivity in the AV (model $H_6$) slightly delays the deformation; although, in the end, it is not able to prevent it. Model $H_7$, calculated with $\sigma=1$, leads to the best results. The cube maintains its profile at $t/\tau_{sc}\simeq 1.7$ (last snapshot in third row in Fig.~\ref{fig:isobaric_cube}) and beyond. Such behavior is in fair agreement with the results of other density-based schemes  \citep{rea10,Wadsley17}~calculated in 2D and 3D respectively. It also matches the 2D calculation in Paper I (first row in Fig. 4 of that work, with $0\le t/\tau_{sc}\le 2$) calculated with the explicit $X_{1,a}$ given by Eq.~(\ref{X1}) with $p=0.7$~(see footnote 7). It is worth noting that model $H_8$, calculated with  $\sigma[0-1]$,  preserves the shape of the square until $t/\tau_{sc}\simeq 1$ and still does at the last snapshot in Fig. \ref{fig:isobaric_cube} at $t/\tau_{sc}=1.7$, although not as well as in model $H_7$. As shown in Fig.~\ref{fig:isobaric_sigma} (upper panel), the algorithm (Eq.~\ref{ramp}) that self-adapts $\sigma$ as a function of the local density contrast works splendidly, increasing the value of $\sigma$ only where it is needed. 

To investigate the dependence of the errors $E_{1,2}$ (Eqs.~\ref{errorE1} and \ref{errorE2}) with respect to the VE estimator and $\sigma$  we calculated the  $L_1$ values (Eq. \ref{L1E_1E_2}) in a rectangular shell with width $0.3$ around the contact discontinuity.

\noindent
 The column $L_1(d)$ of Table~\ref{squaretesttable} is defined by the average error of the absolute displacement of the SPH particles located in the shell, with respect to their initial positions:

\begin{equation}
L_1^{d}(t)=\frac{1}{N}\sum_{b=1}^N \sqrt{(x_b(t)-x_{b,0})^2+(y_b(t)-y_{b,0})^2+(z_b(t)-z_{b,0}^2)}\,,
\end{equation}

\noindent
 where $x$, $y$, and $z$ are the Cartesian coordinates of the particles while $x_0$, $y_0$, and $z_0$ coordinates represent their initial position. Sub-index $b$ runs from 1 up to the $N$ particles contained in the rectangular shell.

The full Lagrangian models $H_1$, $H_2$, $H_5$, and $H_6$, calculated with $\sigma=0$, give the worst results, with $L_1(d) \ge 3\times 10^{-2}$ at $t=1.5$. The minimum value of $L_1(d)$ corresponds to model $H_7$ ($L_1(d)= 1.5\times 10^{-2}$), calculated with $\sigma=1$ and improved VEs. However, model $H_8$ calculated with the self-adaptive $\sigma [0-1]$ and improved VEs, also displays a similarly good behavior, while keeping full Lagrangian consistency ($\sigma\simeq 0$) in a large fraction of the domain (see Fig.~\ref {fig:isobaric_sigma}, upper panel). 

The lower panel in Fig.~\ref{fig:isobaric_sigma} depicts the evolution of the averaged estimator $L_1$ of the partition of the unity (solid lines) and $\left<\Delta r\right>/h$ (dashed lines) for models $H_3$, $H_7$, and $H_8$ in Table~\ref{squaretesttable}.  Comparing $H_3$ to $H_7$ highlights the relevance of the VEs, while comparing $H_7$ to $H_8$ shows the impact of a variable $\sigma$. Is it clear that the partition of the unity is enhanced with the choice $X=m/\rho^0$, nevertheless the behavior of $\left<\Delta r\right>/h$ is less defined. The error $L_1(E_2)$ remains lower for models $H_7$ and $H_8$ than model $H_3$ until $t\simeq 0.2$. Afterwards the error grows to eventually reach a similar value to that of model $H_3$, calculated with $X_a=m_a$. Thus, we expect a degradation in the behavior $\left<\Delta r\right>/h$ with respect the results shown in Figs.~\ref{fig:errorsfunc} and \ref{fig:convergence} whenever the disorder of the particles is high.  In summary, the dominant parameter for the dynamic evolution of the system is $\sigma$, which basically determines the stability of the square. Using $X=m/\rho^0$ improves the partition of unity at any time, but the magnitude $\left<\Delta r\right>/h$ does not show a definite enhancement in this test.   

\begin{table}

        \centering
        \caption{List of simulated models in the Kelvin-Helmholz test. The columns present the following: $\sigma$ is the magnitude defined in Eq.~(\ref{ramp}), $X$ is the estimator connected to the VE choice, $\alpha_{min}$ is the minimum value of the AV parameter, $\rho_1/\rho_2$ is the density ratio, and $L_1(E_1)$ and $L_1(E_2)$ are the averaged $L_1$ values of the partition of unity and the $\left<\Delta r\right>/h$ condition, respectively, at $t=2$.}
        \begin{tabular}{ccccccc} 
                \hline
         \multirow{2}{*}{Model} & \multirow{2}{*}{$\sigma$} & \multirow{2}{*}{$X$} & \multirow{2}{*}{$\alpha_{min}$} & \multirow{2}{*}{$\rho_1/\rho_2$} & $L_1 (E_1)$ & $L_1 (E_2)$\\
         &&&&&$[\times10^{-3}]$&$[\times10^{-3}]$\\
                \hline
                \hline
                $KH_1$ & $0$     & $m$        & 0.05 & 2 & $10.0$ & $4.4$ \\
                $KH_2$ & $[0-1]$ & $m$        & 0.05 & 2 & $9.9$  & $4.3$ \\
                $KH_3$ & $0$     & $m/\rho^0$ & 0.05 & 2 & $2.8$  & $3.7$ \\
                $KH_4$ & $[0-1]$ & $m/\rho^0$ & 0.05 & 2 & $2.9$  & $3.8$ \\
                \hline
                \hline
                $KH_5$ & $[0-1]$ & $m$        & 0.5 & 2 & $10.0$  & $3.5$ \\
                $KH_6$ & $[0-1]$ & $m/\rho^0$ & 0.5 & 2 & $2.2$   & $2.6$ \\
                \hline
                \hline
                $KH_7$ & $[0-1]$ & $m$        & 0.05 & 8 & $18.0$ & $7.7$ \\
                $KH_8$ & $[0-1]$ & $m/\rho^0$ & 0.05 & 8 & $5.8$  & $6.1$ \\
                \hline
                
        \end{tabular}
        \label{KHtesttable}
\end{table}
\begin{figure*}
\includegraphics[angle=0,width=\textwidth]{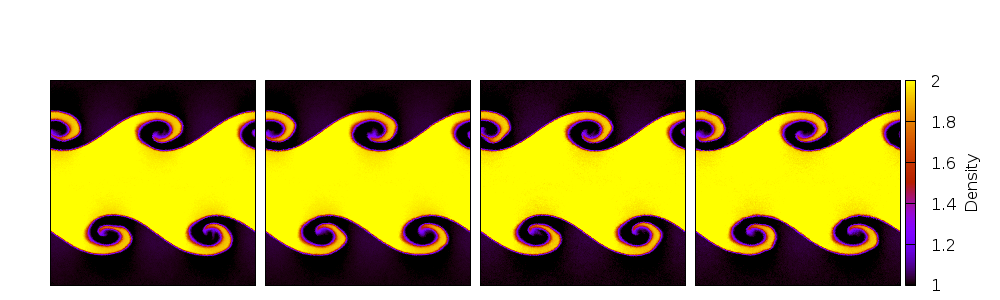}
    \caption{Particles distribution in a thin slice around Z=0 for models $KH_1$ to $KH_4$ at $t=2~ (t_{KH}=1.06)$. The density [1:2] is color-coded.} 
    \label{fig:KH_snapshots}
\end{figure*}

\subsection{Kelvin-Helmholtz instability}
\label{sec-kh}
The correct simulation of the evolution of the contact layer between fluids with different densities is of capital importance for the adequate growth of the Kelvin-Helmholtz (KH) instability. \citet{garciasenz2012} proved that the use of ISPH improves the evaluation of gradients overall (particularly in the contact layer). Provided the density contrast is not very high, it prevents the appearance of tensile instabilities that otherwise suppress the growth of the KH instability. Because SPHYNX uses ISPH by default, all KH simulations presented here show no signs of tensile instability and have growth rates close to the reference calculation by \citet{mcn12} with the code PENCIL. Nevertheless, using a volume element that is better at fulfilling conditions (\ref{errorE1}) and (\ref{errorE2}) should additionally improve the accuracy of ISPH and, as a consequence, obtain a better KH growth rate.

We ran this test in a thin three-dimensional layer of size $[1\times 1\times 0.0625]$ with $4.2\times 10^6$ equal-mass particles, distributed in a ratio 2:1 between the high and low density regions. For the initial setting, we have three stratified layers, being the central layer the high-density one. Each region was generated from a random particle distribution relaxed to a glass-like configuration. The equation of state, initial velocities, and initial pressure are the same as those in Sect. 5.4.1 in Paper I. Times are normalized to the characteristic Kelvin-Helmholtz growth time $t_{KH}$, as defined in \citet{age07}, which results in $t_{KH}=1.06$~for our models. As discussed at the end of Sect.~\ref{subsec:sedov},  we adopted a low value, namely, $\alpha_u = 0.1$, of the AV conductive parameter.

We used two different VEs (the standard $X_a=m_a$ and the enhanced version proposed here $X_a=m_a/\rho^0_a$) and, for each case, we performed two simulations: one fixing $\sigma=0$, hence ensuring fully Lagrangian compatibility, and another one allowing $\sigma$ to vary according to Eq.~(\ref{ramp}). All simulations are summarized in Table~\ref{KHtesttable}.


In Fig.~\ref{fig:KH_snapshots}, we represent the particle distributions for each simulated model (each snapshot corresponds to one model, from $KH_1$ to $KH_4$) at $t=2$. The color represents density. As it can be seen, there are very few differences among all the simulated models. In all cases, the KH billows are able to develop, with small differences among simulations, mostly at the extremes of the billows.

\begin{figure*}
\centering
\includegraphics[width=\textwidth]{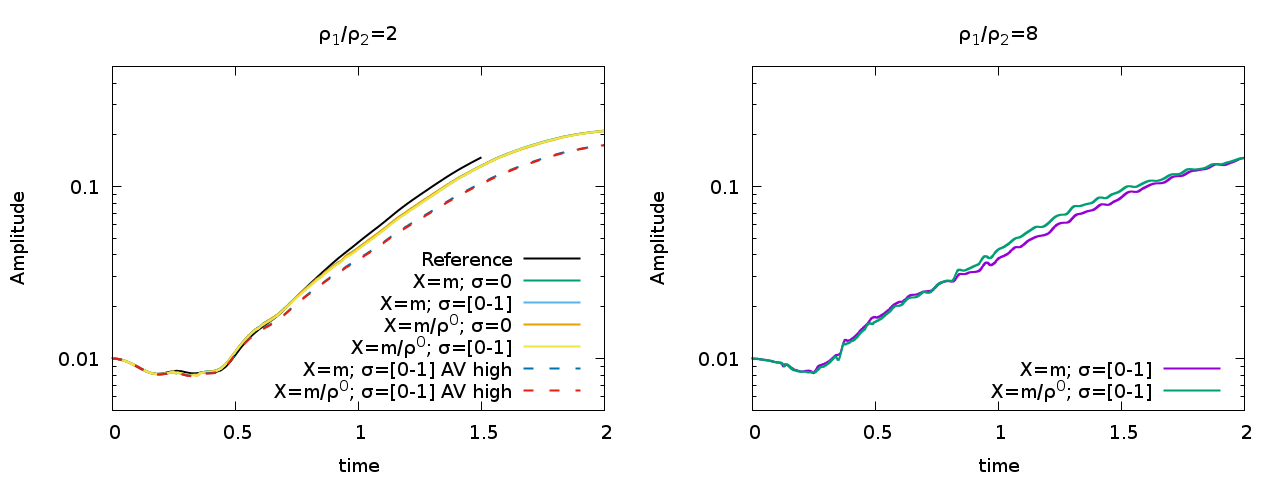}
    \caption{Amplitude growth of the $v_y$ field in the KH instability test for all calculated models. Solid lines correspond to $\alpha_{min}=0.05$, while dashed lines are from the models with increased AV ($\alpha_{min}=0.5$). Those with a density jump by a factor of 2 are compared with the reference PENCIL simulation in the left panel. Right panel shows two simulations with a density jump by a factor of 8 for both choices of the VEs.} 
    \label{fig:KHamplitude}
\end{figure*}

We can also explore the growth of the amplitude mode for the $v_y$ field and compare it with a reference evolution taken from the PENCIL code \citep{mcn12}. Figure~\ref{fig:KHamplitude} (left) shows this evolution and it is clear that there are very little differences among both VEs implementations, in agreement with the results in Paper I. These results are in good match with those by \citet{fro17} and \citet{rosswog2020} for similar resolution and initial conditions, provided that the \citet{cul10} AV trigger is chosen to steer dissipation in the last case.  

\begin{figure*}
\centering
\includegraphics[width=\textwidth]{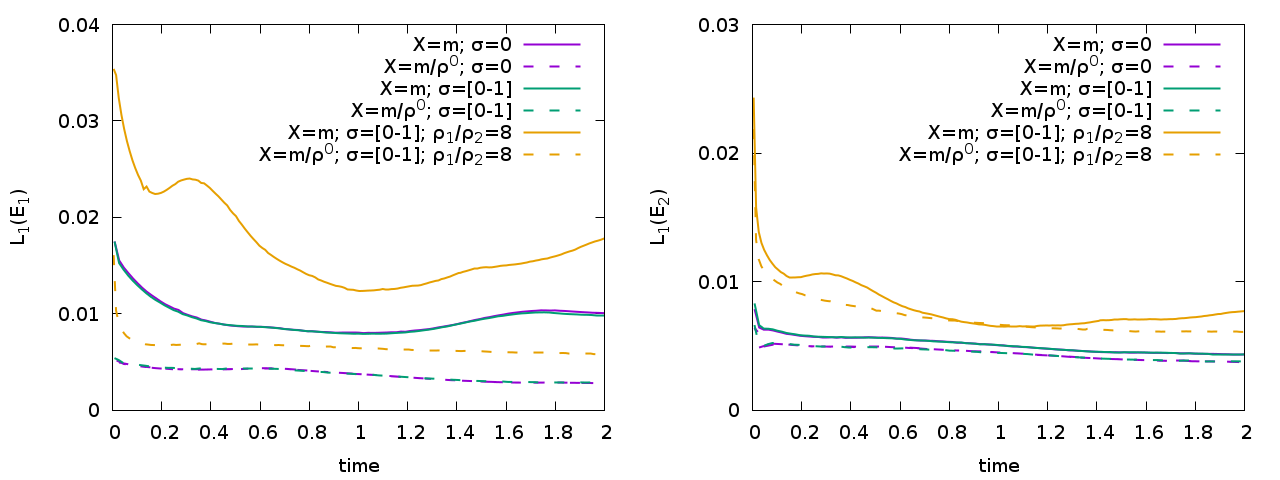}
    \caption{$L_1$ evolution of errors $E_1$ and $E_2$ for the KH models. Solid lines correspond to $X=m$, while dashed lines are those of $X=m/\rho^0$.}
    \label{fig:KHL1}
\end{figure*}

\begin{figure}
\centering
\includegraphics[width=\columnwidth]{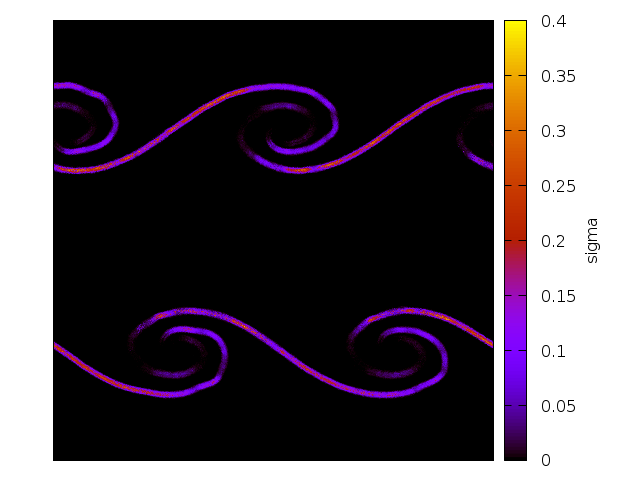}
    \caption{Color map of $\sigma$ (Eq.~\ref{ramp}), in a thin slice around $Z=0$ for model $KH_4$ at $t=2$.}
    \label{fig:sigma}
\end{figure}
Figure~\ref{fig:KHL1} presents the time evolution of the averaged $L_1$ value for errors $E_1$ and $E_2$. To average the $L_1$ values we restricted to those particles that had $\sigma\geq 0.025$. This condition returns all those particles that are found in the inter-phase (see Fig.~\ref{fig:sigma}), which is the region the error comes from and where the accurate evaluation of the VEs is more critical. It is clear that whenever the improved VEs are used (dashed lines in Fig.~\ref{fig:KHL1}), these errors decrease, independently of the scenario being simulated. The last two columns of Table~\ref{KHtesttable} show the numerical value of $L_1$ for all simulated tests at $t=2$, where we can see that the choice $X=m/\rho^0$ reduces both, the $L_1(E_1)$ and the $L_1(E_2)$ errors.

From these results, we can conclude that despite decreasing the errors $E_1$ and $E_2$, the VEs choice has little influence in the growth of the KH instability when the density contrast is moderate. The most important element here is having the IA implementation to calculate gradients, as shown in Paper I. Additional improvement in the conditions (\ref{errorE1}) and (\ref{errorE2}) via using improved VEs is subdominant. The dominant source of error at this level is the artificial viscosity. To test this we performed two additional simulations, $KH_5$ and $KH_6$, with an artificially increased $\alpha_{min}$ in our AV switches. This $\alpha_{min}$ controls the amount of dissipation when the AV switch is off. Therefore, is the minimum dissipation present in the whole system. Our standard value is $\alpha_{min}=0.05$, while for $KH_5$ and $KH_6$ we used $\alpha_{min}=0.5$. To have a reference, traditional AV without switches employ a global value $\alpha_{min}\simeq 1-1.5$. The amplitude growth of the $v_y$ field is also shown in the left panel of Fig.~\ref{fig:KHamplitude} (dashed lines). It is clear that the AV has a major effect in the development of the instability. Still, the VEs have a negligible impact when the IA formalism is used.

Our final test was to explore if the VEs can exert an influence when the density contrast is higher. To do that we decreased the density of the low-density regions by a factor of 4, simply by generating a new relaxed model with four times fewer particles, along with repeated simulations $KH_2$ and $KH_4$, now with this new density jump by a factor of 8 (simulations $KH_7$ and $KH_8$). We present the evolution of the amplitude growth of the $v_y$ field in the right panel of Fig.~\ref{fig:KHamplitude}. In this case, the growth is more irregular and there is a clear difference between the VEs in the linear regime. The improved VEs are able to growth faster than the standard VEs in the initial stages of the development of the KH instability, despite converging later, in the non-linear phase. The IA formalism still does a good job evaluating gradients even with a density jump by a factor 8. Nevertheless, using improved VEs at this density contrast shows to have a noticeable effect.

In summary, if there is a mild density contrast, the IA implementation is good enough to make the choice of VEs sub-dominant. The most relevant parameter in these situations is to keep the dissipation at its lowest possible value, so that random noise is still dissipated, but it doesn't affect the shear between the different fluid layers. In this respect, an improved handling of the AV would be welcomed, such as that presented in \citet{rosswog2020} based on the instantaneous numerical entropy generation rate. If there is a bigger density contrast, the IA formalism can be further improved by using a VE that better fulfills conditions (\ref{errorE1}) and (\ref{errorE2}) along with $\sigma[0-1]$.

\subsection{Wind-cloud collision}

The wind-cloud collision scenario, also called "blob" test \citep{age07} is a challenging test for SPH, involving several pieces of physics such as strong shocks and mixing due to the KH instability, amidst a two-phase medium with a large density contrast. In this test, a spherical cloud of cold gas, initially at rest, is swept by a  low-density stream of gas (the wind) moving supersonically. As a consequence, the cloud deforms and, after a while, it is fragmented and mixed with the background owing to the combined effect of ablation and hydrodynamic instabilities. This scenario has been extensively studied during the last years \citep{age07, rea10, sai13, hop13,fro17}, where the tensile instability appearing at the wind-cloud contact was identified as the main difficulty to overcome. 

We sought to check the ability of our numerical scheme to handle this test. The initial conditions were those described in Paper I, except that the calculation is now carried out in 3D with approximately $N_w=11.23\times10^6$ and $N_c=1.23\times10^5$ particles for the wind and the cloud, respectively. The initial particle distribution for both was that of a stable glass-like configuration. The box has a size \{$0.25\times 0.25\times 1$\} with periodic boundary conditions. The high-density cloud is initially located at $(0.125,0.125,0.125)$ with a radius $R=1/40$ and a density $\rho_c=10$, ten times bigger than the surrounding wind. The internal energy of the wind and the cloud are $u_w=3/2$ and $u_c=3/20$, respectively, so that both regions are in pressure equilibrium. The wind has an initial velocity $(2.7,0,0)$. {The characteristic KH growth time used in this test to normalize the time is $\tau_{KH}=0.0937$.}
\begin{table}
        \centering
        \caption{Simulated models in the wind-cloud test. The columns present the following: $\sigma$ is the magnitude defined in Eq.~(\ref{ramp}), $X$ is the estimator connected to the VE choice, and $L_1(E_1)$ and $L_1(E_2)$ are the averaged $L_1$ values of the partition of unity and the $\left<\Delta r\right>/h$ condition, respectively, at $t/t_{KH}=3$.}
        
        \begin{tabular}{ccccc} 
                \hline
                \multirow{2}{*}{Model} & \multirow{2}{*}{$\sigma$} & \multirow{2}{*}{$X$} & $L_1(E_1)$ &$L_1(E_2)$\\
                &&&$[\times10^{-2}]$&$[\times10^{-2}]$\\
                \hline
                \hline
                 $W_1$& 0 & $m$ & 3.84 & 2.17\\
                 $W_2$& 1 & $m$ & 2.79 & 1.43\\
                 $W_3$& $[0-1]$ & $m$ & 2.60 & 1.35\\
                 $W_4$& 0 & $m/\rho^0$ & 1.39 & 1.76\\
                 $W_5$& 1 & $m/\rho^0$ & 0.93 & 1.12\\
                 $W_6$& $[0-1]$ & $m/\rho^0$ & 0.92 & 1.12\\
                \hline
        \end{tabular}
        \label{tab:blobtests}
\end{table}

\begin{figure}
\centering
\includegraphics[width=\columnwidth]{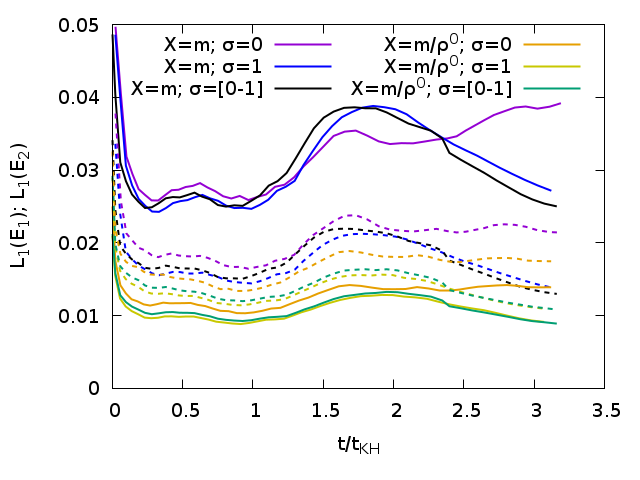}
    \caption{$L_1$ values for $E_1$ (solid lines) and $E_2$ (dashed lines) in the wind-cloud collision tests. Using the improved VEs reduces both $L_1$ values for the whole simulation.}
    \label{fig:L1_windblob}
\end{figure}

\begin{figure}
\centering
\includegraphics[width=\columnwidth]{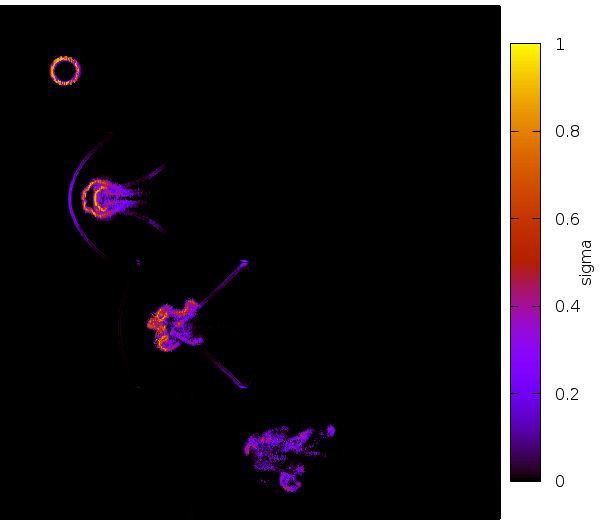}
    \caption{Particle distribution of the wind-cloud test in a thin slice around $Y=0$ for model $W_6$ at $t/t_{KH}=0$, 1, 2, and 3. The parameter $\sigma$ is color-\allowbreak coded.}
    \label{fig:sigma_windblob}
\end{figure}

In Table~\ref{tab:blobtests} we present the $L_1$ values of errors $E_1$ and $E_2$ for the simulated models. Both errors are clearly lower when the improved VEs are used, independently of the value of $\sigma$. This trend is constant during the whole evolution of the simulation as it can be seen in Fig.~\ref{fig:L1_windblob}.  In particular, it is clear that the models with $\sigma=0$ lead to the largest $L_1$ errors in the long run, proving that the pure Lagrangian scheme is the most sensitive to tensile instability with such large density gradients. As in the case of the KH test, we make use of a variable  $\sigma$ to track the $L_1$ error generated by the particles in the inter-phase. In Fig.~\ref{fig:sigma_windblob} we show a series of snapshots where the color represents the value of $\sigma$ and proves the capabilities of our algorithm to track density jumps. 

\begin{figure}
\centering
\includegraphics[width=\columnwidth]{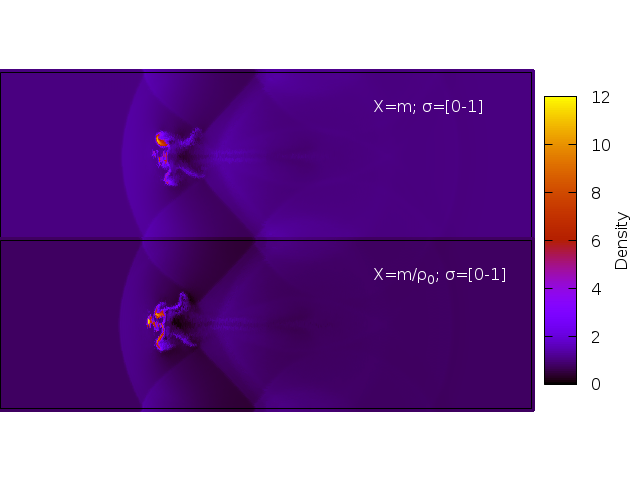}
    \caption{Particle distribution of the wind-cloud test in a thin slice around $Y=0$ for models $W_3$ and $W_6$ at $t_{KH}=1.5$. Density is color-coded.}
    \label{fig:windblob}
\end{figure}

\begin{figure}
\centering
\includegraphics[width=\columnwidth]{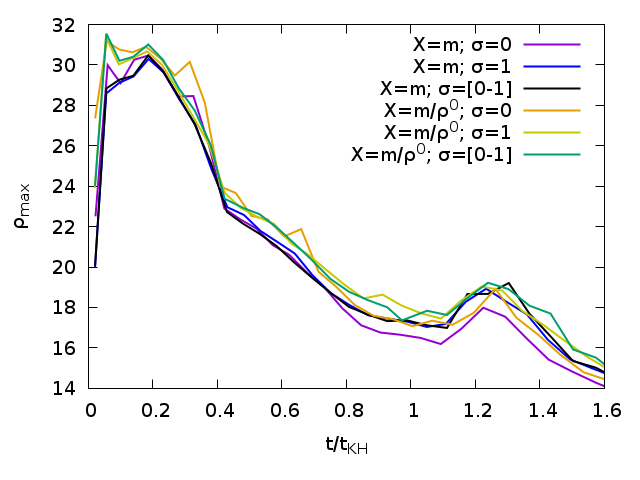}
    \caption{Evolution of the maximum density in the cloud at $t/t_{KH}\le 1.6$ for all models in Table~\ref{tab:blobtests}.}
    \label{fig:romax}
\end{figure}

 The density color map at $t/t_{KH}=1.5$ is shown in Fig.~\ref{fig:windblob} for both choices of VEs.  We see here that, independently of the VE choice, mixing is taking place, which will eventually destroy the cloud -- albeit more efficiently if $\sigma\ne 0$. Figure~\ref{fig:romax} shows the maximum density attained in the cloud during the first stages of the interaction. Because of the strong impact of the wind, a shock-wave is born, which moves through the dense cloud and compresses its matter. The density jumps from the original value of $\rho_0=10$ and tends toward the limiting value $\rho=4 \rho_0$, which is characteristic of a strong shock in a gas with $\gamma=5/3$. All models show a similar behavior but the profile from models calculated with $X=m/\rho^0$, give in general larger values than models calculated with  $X=m$.       
 
\begin{figure}
\centering
\includegraphics[width=\columnwidth]{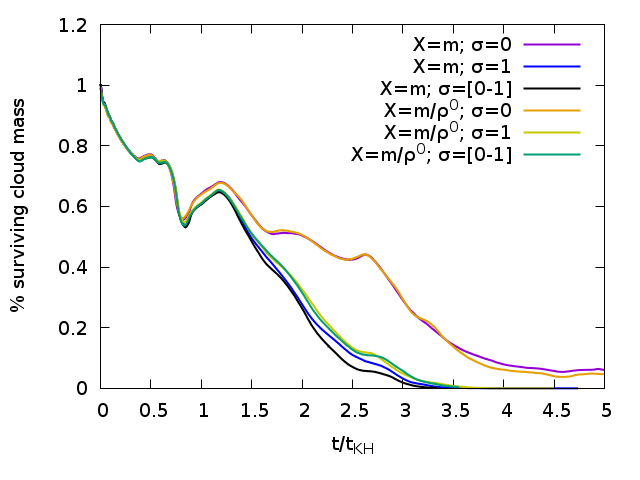}
    \caption{Percent of surviving cloud in function of $t/t_{KH}$ for the wind-cloud collision tests.}
    \label{fig:survivingcloud}
\end{figure}

In order to characterize the destruction of the high-density cloud, we tracked the percent of surviving cloud mass (criteria: $\rho\geq 0.64\rho_c$ and $u\leq 0.9 u_w$) in function of $t/t_{KH}$ in Fig.~\ref{fig:survivingcloud}. From this result, we can see that the choice of the VEs is less relevant than using a crossed formulation of the equations (i.e., variable $\sigma$ or $\sigma=1$). Indeed, the cloud is mostly destroyed in all simulations, and this is mainly due to the integral formulation of the SPH equations combined with the addition of the heat transfer term in the AV equation. Nevertheless, the surviving fraction of the cloud decreases faster and to lower values -- provided we do not use $\sigma=0$, independently of the VE choice. The slight delay in the evolution of the surviving fraction with  $X=m/\rho^0$ simply reflects the fact that the ensuing VEs better track the density peak at the forward shock moving through the cloud. Such a density enhancement results in the slight delay in crossing the above threshold condition $\rho\geq 0.64\rho_c$.  The disintegration rate of the cloud depicted in Fig.~\ref{fig:survivingcloud} when $\sigma\ne 0$ is in close agreement with that obtained with the CRKSPH scheme by \citet{fro17}.

 In summary, when large density contrasts are present, better accuracy is achieved by the proposed generalized volume elements, while the suppression of the tensile instability is handled via a surgical departure of the Lagrangian scheme, using a variable $\sigma$.

\subsection{Sedov explosion}
\label{subsec:sedov}

We simulated this test in a three-dimensional square box of side $L=1$ and total number of particles $N=100^3$ arranged in a glass-like configuration. The mass of all the particles was the same, resulting in an homogeneous initial density profile with \mbox{$\rho=1\pm 0.005$}. The explosion was initiated at the center of the box by depositing an amount of energy $\Delta U=1$. That energy was spread following a Gaussian profile with characteristic width $\delta=0.1$.

\begin{table}\centering
        \caption{Summary of all calculated Sedov tests stating the values of $\sigma$, heat transport constant in the AV, VE choice, density peak, and $L_1$ errors at $t=0.09$}
        \begin{tabular}{ccccccccc} 
                \hline
         \multirow{2}{*}{Model} & \multirow{2}{*}{$\sigma$} & \multirow{2}{*}{$\alpha_u$} & \multirow{2}{*}{$X$}& \multirow{2}{*}{$\rho_{max}$} & $L_1(E_1)$&$L_1 (E_2)$ \\
         &&&&&$[\times10^{-2}]$&$[\times10^{-2}]$\\
             \hline
                \hline
                $S_1$ & $0$ & $0.1$ & $m$ &$3.34$&$3.10$&$2.20$\\
                $S_2$ & $0$ & $0.5$ & $m$ &$3.16$&$2.85$&$1.96$\\
                $S_3$& $1$ & $0.1$ & $m$ &$3.33$&$3.05$&$2.20$\\
                $S_4$& $[0-1]$ & $0.1$ & $m$ &$3.35$&$3.10$&$2.20$\\
                $S_5$& $0$&$0.1$&$m/\rho^0$ &$3.71$&$1.53$&$1.72$\\
                $S_6$& $0$ & $0.5$ & $m/\rho^0$&$3.47$& $1.33$&$1.52$\\
                $S_7$& $1$ & $0.1$ & $m/\rho^0$&$3.70$& $1.45$&$1.67$\\
                $S_8$& $[0-1]$ & $0.1$ & $m/\rho^0$&$3.71$& $1.45$&$1.63$\\
                \hline
        \end{tabular}
        \label{sedovtable}
\end{table}

\begin{figure}
\centering
\includegraphics[angle=0,width=\columnwidth]{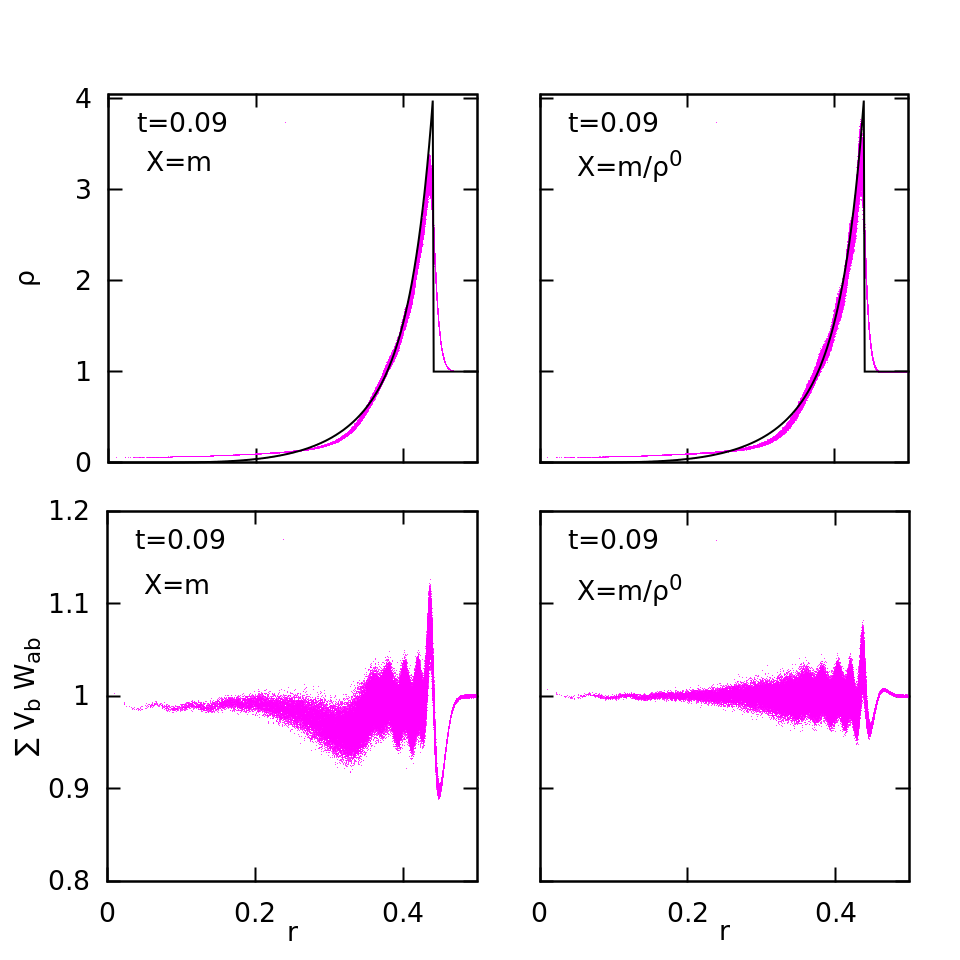}
    \caption{Sedov explosion for models $S_4$ and $S_8$ in Table (\ref{sedovtable}). Top row: density profiles at $t= 0.09$ calculated with $X_a=m_a$ (model $S_4$, standard VEs) and $X_a=m_a/\rho_a^0$ (model $S_8$, improved VEs). The black line is the analytical solution. Note that the spherical symmetry is well preserved in both cases and that the density peak is very well reproduced with the improved VEs.
    Bottom row: Same but with the partition of unit. All particles are represented in the plots. }
    \label{fig:sedov_1}
\end{figure}

\begin{figure}
\centering
\includegraphics[angle=0,width=\columnwidth]{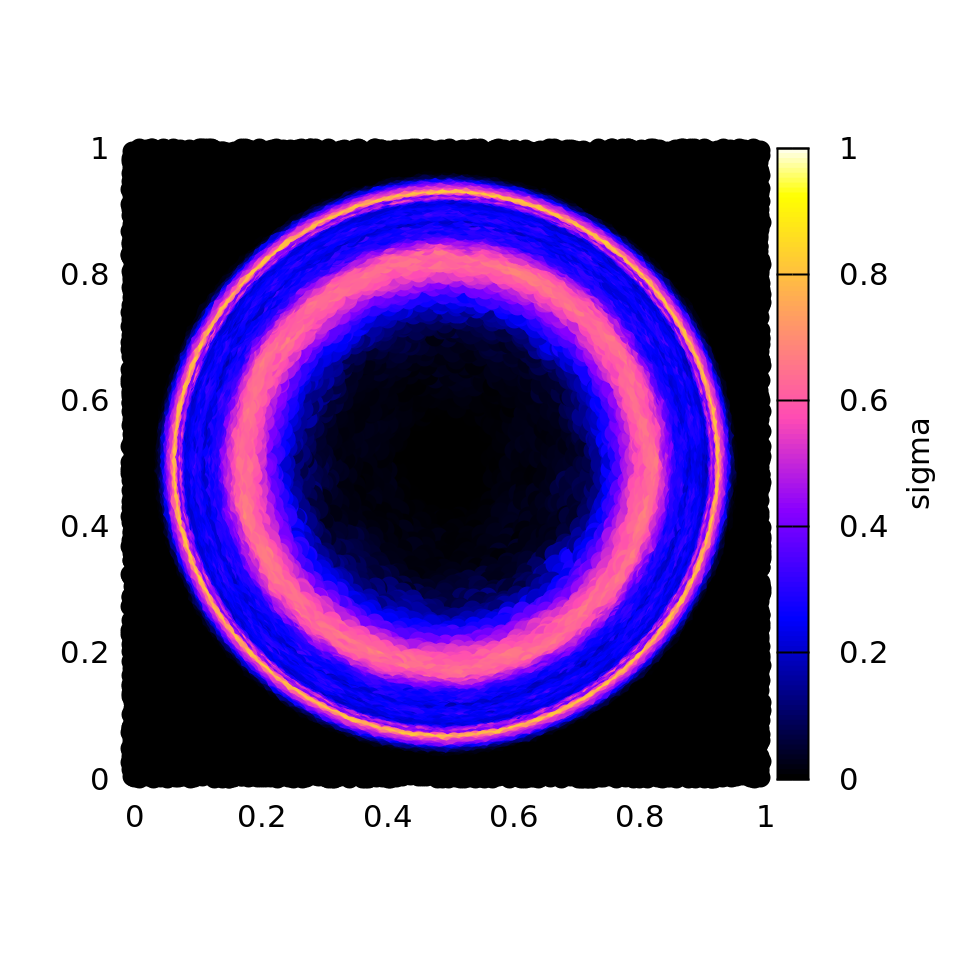}
    \caption{Color map slice  around $Z=0$ showing the value of the averaged $\sigma$-parameter~defined in Eq.~(\ref{ramp}), for model $S_8$ in Table~\ref{sedovtable} at $t=0.09$.}
    \label{fig:sedov_2}
\end{figure}

\begin{figure}
\centering
\includegraphics[angle=0,width=\columnwidth]{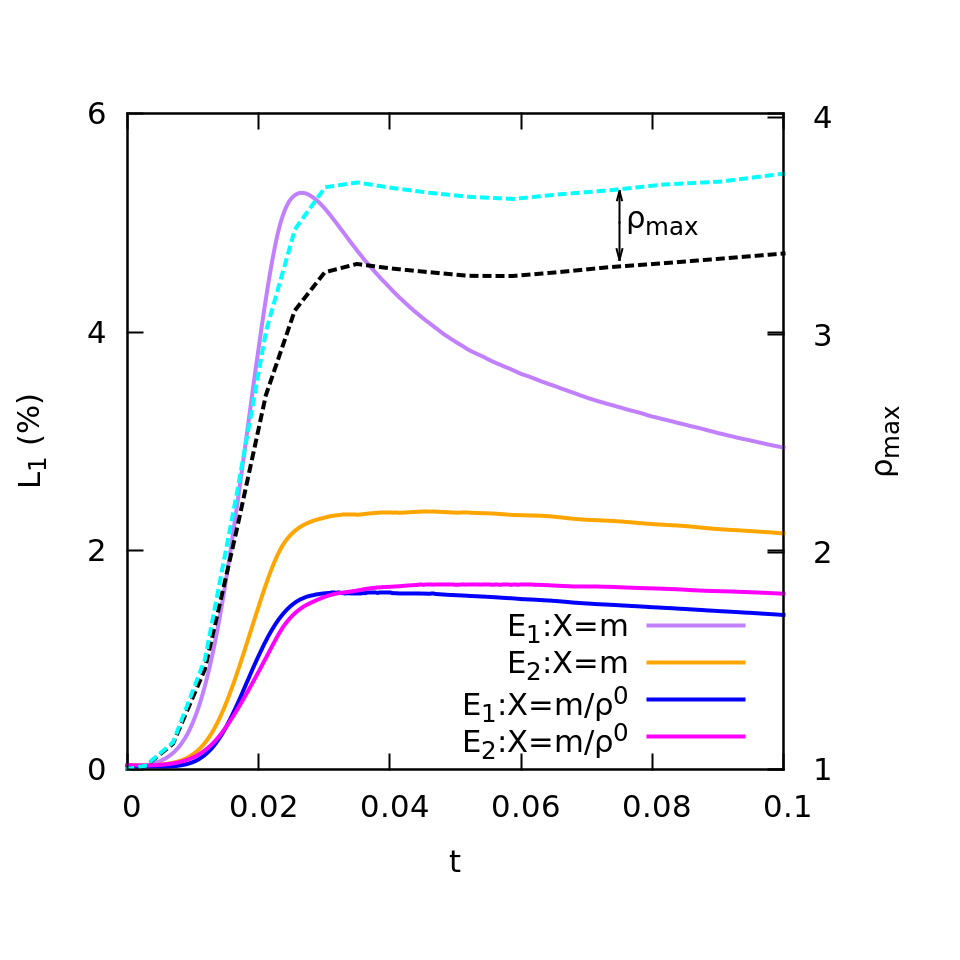}
    \caption{Sedov explosion. Evolution of the averaged $L_1$ error, (Eq.~\ref{L1E_1E_2}), in the shocked region corresponding to models $S_4,~(X_a=m_a)$, and $S_8,~(X_a=m_a/\rho^0_a)$, in Table \ref{sedovtable}. We show both the partition of unity and the normalized $\left<\Delta r\right>/h$ condition, for the estimator choices: $X_a=m_a$ (purple and orange lines) and $X_a=m_a/\rho^0_a$ (blue and magenta lines). Also shown is the evolution of the maximum density for the two volume estimators (dashed lines, black is for $S_4$ and light-blue for $S_8$) during the explosion. } 
    \label{fig:sedov_3}
\end{figure}

 To investigate the dependence of the errors $E_{1,2}$ (Eqs.~\ref{errorE1} and \ref{errorE2}) with respect the VEs estimator choice and $\sigma$, we calculated the $L_1$ errors 
 in the shocked region\footnote{Defined as the volume with specific internal energy $ u(t)\ge 0.1$} with Eq.~(\ref{L1E_1E_2}).

The outcome of the simulations is summarized in Table~\ref{sedovtable} and Figs.~\ref{fig:sedov_1}, \ref{fig:sedov_2}, and \ref{fig:sedov_3}. The best results were obtained in those models which incorporate the improved VEs. Unlike in the hydrostatic square test, the particular value adopted for the magnitude $\sigma$ plays a secondary role in the Sedov test. Figure~\ref{fig:sedov_1} shows the profiles of density at t=0.09 for models $S_4$ and $S_8$. In both cases, the self-adaptive algorithm to estimate $\sigma$ was active. The best match with the analytic profile (black line) was obtained with $X_a=m_a/\rho^0_a$, not only in the peak values, which were better reproduced, but also in the width of the shock front. The relative errors of the maximum density with respect to the analytical value at $t=0.09$ are $\simeq 16\%$ and $\simeq 7\%$ for the standard ($S_4$) and improved VEs ($S_8$), respectively. Figure~\ref{fig:sedov_2} shows the distribution of the averaged $\sigma$ in a thin slice along the system for model $S_8$. As it can be seen, the value of $\sigma$ self-adapts so that it approaches to 1 across the shock front, goes down to $\simeq 0.5$ in the post-shock region and vanishes in the central region. 

The profile of the normalization function $\sum_b V_b W_{ab}(h_a)$ at $t=0.09$ is also shown in Fig.~\ref{fig:sedov_1}, bottom row. As it can be seen, the choice $X_a=m_a/\rho^0_a$ substantially improves the partition of the unity with respect the standard choice $X_a=m_a$. Such assessment is quantitatively confirmed by the temporal evolution of the error estimators $L_1 (E_1, E_2)$ shown in Fig.~\ref{fig:sedov_3}. The enhancement is especially good in the case of the partition of unity, although less pronounced in the $\left<\Delta r\right>/h$ condition. Therefore, we conclude that combining ISPH with the VEs obtained with \mbox{$X_a=m_a/\rho^0_a$} reduces both errors and improves the simulations. 

The impact of increasing the conductive term in the AV was also analyzed. Raising the parameter $\alpha_u$ from our default choice, $\alpha_u=0.1$ to \mbox{$\alpha_u=0.5$}, substantially reduces the density peak from \mbox{$\rho_{max}(t=0.09)=3.34$} in model $S_1$ to \mbox{$\rho_{max}(t=0.09)=3.16$} in model $S_2$. In light of these results, it is advisable to choose a not too high value for $\alpha_u$. 

\subsection{Choice of \texorpdfstring{$\sigma$}{[sigma]} and entropy evolution in a shock}
\label{subsubsec:entropy}

In light of the results of the tests presented here, the question of why we would need a variable $\sigma$ at all might arise. Obviously, the case $\sigma=0$ would be the most desirable one, as it is fully Lagrangian-compatible. Nevertheless, it is clear that the crossed $\sigma=1$ scheme leads to a more efficient removal of the tensile instability, but the price to pay is the lack of a complete compatibility between the Euler-Lagrange formulation and the standard density equation (Eq.~\ref{density_1}). Such a weakness with regard to the crossed scheme is highlighted in the following test, which considers the behavior of entropy in weak and strong shocks. We want to prove that the entropy does not evolve as well when $\sigma\ne 0$. We additionally show that the new volume elements have a positive impact on the evolution of entropy.

Our numerical experiment is the same as in the Sedov test above but with an additional case where the initial energy of the explosion has been reduced in a factor of ten to simulate the case of a weak explosion. During the evolution of the explosion, the contribution to the entropy of the gas is separated in two components: that arising from pressure forces $\Delta s_P$ and that from the artificial viscosity $\Delta s_{AV}$. Ideally, $\Delta s_P\simeq 0$ because the dissipation is basically driven by the AV.       

The second law of thermodynamics states:

\begin{equation}
T ds = du - \frac{P}{\rho^2}d\rho\,,
\label{entropy_1}
\end{equation}

\noindent
where $T, s, u, P$ are temperature, specific values of entropy and internal energy, and pressure. 
For an ideal gas with specific heat at constant volume, $c_v$, we have:

\begin{equation}
  \frac{ds}{c_v}=\frac{du}{u}- (\gamma-1)~\frac{d\rho}{\rho}\,,
\label{entropy_2}
\end{equation}

\noindent with $u=c_v~T$. Thus, integrating on both sides:

\begin{equation}
 \frac{\Delta s}{c_v}=\ln\left(\frac{u}{u_0}\right)-(\gamma-1)\ln\left(\frac{\rho}{\rho_0}\right)\,,
\label{entropy_3}
\end{equation}

\noindent  where  $\rho_0, u_0$~are the initial values of density and internal energy. 
To obtain the evolution of entropy of the ideal gas, without the AV contribution, we first calculate the increment of entropy $(\Delta s)_{AV}/c_v$ due to the AV. Subtracting $(\Delta s)_{AV}/c_v$ from (\ref{entropy_3}) gives the ideal gas contribution, which should remain negligible during an adiabatic evolution.

The AV contribution to the entropy in a process involving an energy/heat variation $dQ$~is:

\begin{equation}
\frac{dQ}{T}= (ds)_{AV}\rightarrow \frac{c_v~dQ}{u}= (ds)_{AV} \rightarrow \frac{ds_{Av}}{c_v}=\frac{dQ}{u}\,,
\label{entropy_4}
\end{equation}

\noindent
which, integrating on both sides, leads to:

\begin{equation}
\frac{(\Delta s)_{AV}}{c_v}=\int \frac{dQ}{u}=\int_0^{t} \frac{1}{u}\left(\frac{du}{dt}\right)_{AV} dt\,,
\label{entropy_5}
\end{equation}

\noindent
where $\left(\frac{du}{dt}\right)_{AV}$ is directly obtained from the SPH code. The variation of the entropy of the gas, due to pressure forces $(\Delta s)_{P}/c_v$, and excluding the AV contribution, is determined subtracting Eq. (\ref{entropy_5}) from Eq. (\ref{entropy_3}):

\begin{equation}
\frac{(\Delta s)_{P}}{c_v}= \frac{\Delta s}{c_v}-\frac{(\Delta s)_{AV}}{c_v}\,.
\label{entropy_6}
\end{equation}

In absence of heat transport and providing that there are neither entropy sources (e.g., nuclear reactions) nor entropy sinks, $(\Delta s)_{P}/c_v\simeq 0$.

\begin{figure}
\centering
\includegraphics[angle=0,width=\columnwidth]{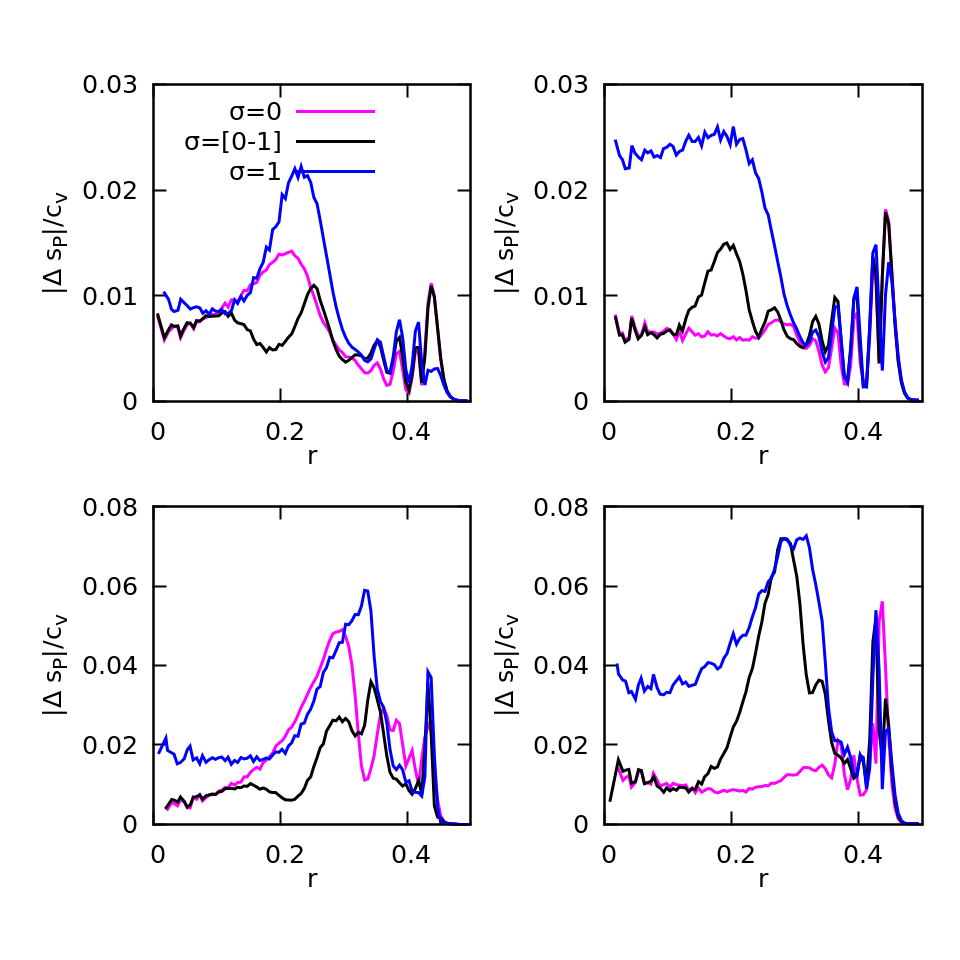}
    \caption{Radial profile of the entropy variation by pressure forces, $\vert\Delta s_P\vert/c_v$, for weak (upper row, at $t=0.18$) and strong (lower row, at $t=0.08$) point-like explosions and different $\sigma$ choices. The first column is for estimator $X_a=m_a$ and the second for $X_a=m_a/\rho^0_a$. } 
    \label{fig:entropy_1}
\end{figure}

In Fig.~\ref{fig:entropy_1}, we show the radial profile of $\vert\Delta s_{P}\vert/c_v$ for the different calculated models. First of all, we note that there is a spurious variation of entropy in all cases. Nevertheless, the combination  $(X_a=m_a/\rho^0_a; \sigma=0)$~shown in the fourth panel leads to the best behavior with regard to the entropy in the shocked region (magenta lines). Not surprisingly, the worst cases are for $\sigma=1$ (blue lines). These results clearly prove that Lagrangian compatibility is of utmost relevance for keeping the entropy variations at low levels in adiabatic flows. Furthermore, upgrading the partition of unity with a Lagrangian-compatible estimator, such as  $X_a=m_a/\rho^0_a$, substantially improves the results. The mixed option with $\sigma=[0-1]$ (black lines) is, for most of the cases, closer to $\sigma=0$~or in between both curves, as expected. It therefore represents a balanced solution between Lagrangian-compatibility and  the suppression of the tensile instability. 

\begin{figure}
\centering
\includegraphics[angle=0,width=\columnwidth]{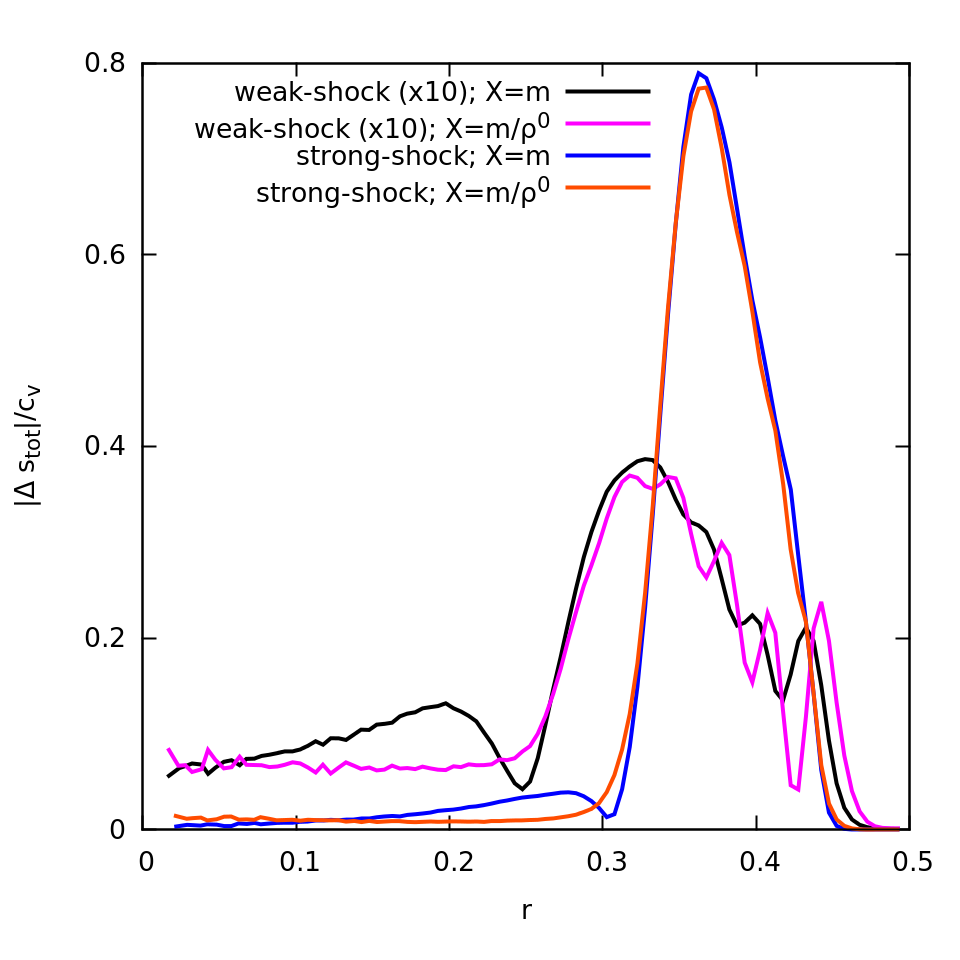}
    \caption{Radial distribution function of total entropy, $\vert\Delta s_{tot}\vert/c_v$, released in weak or strong shocks with $\sigma=0$ and the two VEs considered in this work. The entropy in the weak shock cases has been multiplied by 10. } 
    \label{fig:entropy_2}
\end{figure}

The profiles of the total entropy variation is shown in Fig.~\ref{fig:entropy_2}. The strong shock dissipates much more entropy than the weak. Thus, the spurious generation of entropy in weak shocks, as shown in the upper panels of Fig.~\ref{fig:entropy_1}, is comparatively more acute. 

The results of this test support the idea that working with an adaptive $\sigma$ is a useful option that allows us to suppress the tensile instability across shocks and contact discontinuities, while reducing the unwanted growth of entropy in the post-shock regions. 

\section{Conclusion}

\begin{table*}[ht!]
       \centering
       \caption{Summary of the features of the approach presented in this work within the context of the works on which it is based. The first column shows the method, where IAD stands for the integral approach to derivatives, which is at the core of our proposed ISPH; GVE stands for generalized volume elements, and $\sigma$ is the methodology presented in this work that is used to obtain a crossed version of the momentum and energy equations only where it is needed to suppress the tensile instability. The second and third columns give a qualitative idea of the associated accuracy when evaluating gradients and when then tensile instability is suppressed. The fourth and fifth columns state if the method is fully conservative and Lagrangian compatible. The last column shows the associated reference. We note that the first row corresponds to the ncISPH scheme mentioned at the end of Sect.~\ref{sec:isph}, whereas the other three are conservative ISPH schemes.}
       \resizebox{\textwidth}{!}{%
        \begin{tabular}{cccccc} 
                \hline
               Method & Gradient accuracy & Tensile Instability suppression & Conservative$^{\dgs a}$ & Lagrangian compatibility$^{\dgs b}$ & Reference\\
               \hline
               \hline
                IAD & Very high & Small and moderate density jumps & No & No & \citet{garciasenz2012}\\
                IAD$_0$ & High & Small and moderate density jumps &  Yes & Yes & \citet{cabezon2012}\\
                IAD$_0$ + GVE (Explicit) & Very high & Large density jumps & Yes & No & \citet{cabezon2017}\\
               IAD$_0$ + GVE (Implicit) +~$\sigma$ & Very high & Large density jumps & Yes & Mostly yes & This work\\
               \hline
       \end{tabular}}
       \footnotesize{
       $^{\dgs a}$Linear momentum, angular momentum and energy; $^{\dgs b}$Leading to an optimal handling of entropy }
        \label{tablefinal}
\end{table*}

 The main goal of this work is to devise and check an SPH scheme that fulfils as many basic physical and numerical constraints as possible, with regard to the demands of modern hydrodynamic codes. The former include conservation of mass, linear and angular momentum, energy, and entropy. The latter requires a good depiction of gradients and shocks besides a correct numerical representation of basic kernel properties, as for example, the kernel normalization. The proposed scheme strongly relies in the ISPH method plus enhanced volume elements described in \citet{cabezon2017}, but with an improved Lagrangian compatibility and a better treatment of  sharp density gradients. 

To this end, here we introduce an easy scheme to improve the partition of the unity in SPH, with a special emphasis on the connections between the partition of unity and the accuracy in estimating the gradients. By combining analytical reasoning with simple 1D toy models and full 3D simulations, we have shown that improving the constraint $\sum_b V_b W_{ab}(h_a)=1$ automatically leads to an enhancement of the condition $\left<\Delta r\right>=\sum_b V_b ({\bf r}_b-{\bf r}_a)W_{ab}(h_a)=0$. When gradients are approached with an integral expression (the ISPH scheme via Eq.~\ref{matrix}), the fulfillment of the $\left<\Delta r\right>$ constraint is shown to be a sufficient condition to perform exact linear interpolations. 

 One of the novelties of this work, is the improvement of the partition of the unity, without leaving the Lagrangian formulation of ISPH, so that the enhancement is totally compatible with the inclusion of grad-h terms in the momentum and energy equations. To do that the volume elements, $V_a$, have been re-defined so that \mbox{$V_a=X_a/k_a$} with \mbox{$X_a=m_a/\rho^0_a$;  $k_a=\sum_b X_b W_{ab}(h_a)$} and \mbox{$\rho^0_a=\sum_b m_b W_{ab}(h_a)$}. That particular choice of $V_a$ results in the new set of ISPH equations described in Appendix~\ref{sec:Appendix}. These ISPH equations are not only  Lagrangian-compatible, displaying perfect conservation properties, but they manifestly improve the partition of the unity which translates into a better estimation of gradients. The computational cost of including these improved VEs, mainly due to the implicit search of the coupled $h_a$ and $\rho^0_a$ magnitudes with a Newton-Raphson algorithm, is subdominant. The behavior of the new ISPH scheme has been checked with a number of standard 3D tests that involve large density jumps and contact discontinuities.

A second novelty in the present work has to do with the correct handling of abrupt density jumps. The numerical handling of fluid regions with large density contrasts usually requires us to abandon the Lagrangian formulation of the SPH equations in favor of other schemes. We propose here a self-adaptive scheme, steered by a single parameter of $0\le \sigma\le 1$ (see Eqs. \ref{mom_1}, \ref{energy_1}, and \ref{ramp}), which selectively chooses the optimal integration scheme. When $\sigma\simeq 0,$ the Lagrangian-compatible scheme is recovered, while $\sigma\simeq 1$ is more adequate to handle large gradients.

 The proposed scheme with $\sigma=[0-1]$~gets rid of the tensile instability around contact discontinuities, as it is the closest possible to the compatibility with the Lagrangian formalism because the fraction of particles with a significant $\sigma$ value is usually low. We can therefore expect to obtain a better depiction of the entropy evolution in shocks than in completely crossed schemes (i.e., with $\sigma =1$). According to Eq.~(\ref{ramp}), a value of $\sigma \le 0.1$ is taken in those fluid regions with $\frac{\Delta\rho}{\bar \rho}\le 22\%$, meaning that acoustic waves and weak shock-waves can be handled with, or bordering,  the uncrossed option $\sigma \simeq 0$. Such moderate density ratios over the kernel domain are also found in self-gravitational structures in equilibrium and in subsonic turbulence experiments.

 Table~\ref{tablefinal} presents a summary of the above-mentioned contributions and contrasts them with our previous works, which sets the current extension into context. Following \citet{garciasenz2012}; \citet{cabezon2012},  our chosen method  was ISPH, which uses the integral approach to the derivatives, IAD$_0$, to implement the equations. Further developments were bound to the inclusion of generalized volume elements that improved the partition of the unity.

The results of the hydrodynamic tests unambiguously support the conclusions drawn by the analytical arguments and simple toy models described in Section \ref{sec:static}. In particular, we find that the proposed volume elements improve both the partition of the unity and the $\left<\Delta r\right>$ condition in all studied cases. Nevertheless, the level of enhancement in the hydrodynamic simulations is lower than that promised by the toy model experiments, especially those concerning the $\left<\Delta r\right>$ constraint. Such a degradation is attributable to the larger particle disorder in the real three-dimensional simulations, but still the results with the improved VEs are valuable. On another note, the novel self-adaptive $\sigma$ scheme works very well, providing much better results in the hydrostatic and wind-cloud tests than the $\sigma=0$ calculation. The results of the Kelvin-Helmholtz test suggest that the artificial viscosity algorithm plays a central role here, with the VEs and $\sigma$ choice being subdominant. Finally, the Sedov test clearly shows the best performance of the proposed VEs, which allow us to reproduce the correct density jump across the shock, even in three-dimensional calculations with moderate number of particles. Interestingly, the improved evolution of entropy in shocked regions when the $\sigma=[0-1]$ scheme is used (see Sect.~\ref{subsubsec:entropy}) reinforces the recent proposal by \citet{rosswog2020}, who suggested using the entropy as a useful variable for building physically motivated AV triggers in SPH.

As an immediate prospect, we plan to use our improved ISPH code to numerically reproduce the isothermal as well as sub- and supersonic turbulence. Works to implement the ISPH scheme to handle magneto-hydrodynamics effects are also underway.

\section*{Acknowledgments}
We thank the anonymous referee for insightful comments and suggestions that helped to greatly improve this manuscript.  This work has been supported by the MINECO Spanish project PID2020-117252GB-100, by the Swiss Platform for Advanced Scientific Computing (PASC) project SPH-EXA: Optimizing Smooth Particle Hydrodynamics for Exascale Computing (R.C. and D.G.). The authors acknowledge the support of sciCORE (http://scicore.unibas.ch/) scientific computing core facility at University of Basel, where part of these calculations were performed.

\bibliographystyle{aa}

\begin{appendix}
\section{SPH and ISPH formalisms with generalized VE}
\label{sec:Appendix}

According to the Euler-Lagrange formulation of SPH (\citet{spr10} and references therein) the movement equations are expressed as: 

\begin{equation}
    m_a \ddot{\mathbf r}_a=-\sum_b m_b\frac{P_b}{\rho_b^2}\frac{\partial\rho_b}{\partial {\mathbf r}_a} = \sum_b P_b \frac{\partial V_b}{{\partial \mathbf r}_a}
    \,,
    \label{app_1}
\end{equation}
\noindent
where $V_b$ is the characteristic volume occupied by the particle, and \mbox{$\rho_b=m_b/V_b$}. The derivative on the RHS embodies the effect of $h$-gradients. The spatial part of the derivative can be performed in the standard way, $\nabla W_{ab}$ or, better, with the integral approach given by Eq. (\ref{stdiad}) 

We first calculate the value of $\partial V_b/\partial {\mathbf r}_a$ in Eq.~(\ref{app_1}),

\begin{equation}
    \frac{\partial V_b}{{\partial \mathbf r}_a} = \nabla_a V_b + \frac{\partial V_b}{\partial h_b}\frac{\partial h_b}{\partial {\mathbf r}_a}
    \,.
    \label{app_2}
\end{equation}

 The grad-h part of equation above can be estimated differentiating the constraint $h_b^3 V_b^{-1}= C$ with respect ${\mathbf r}_a$ which, after some algebra, gives:
 
 \begin{equation}
     \frac{\partial V_b}{\partial h_b}~\frac{\partial h_b}{\partial {\mathbf r}_a}\left[1-\frac{3V_b}{h_b}\left(\frac{\partial V_b}{\partial h_b}\right)^{-1}\right]=-\nabla_a V_b
     \,,
 \end{equation}
 
 \noindent when combined with expression (\ref{app_2}) gives:
 
 \begin{equation}
     \frac{\partial V_b}{{\partial \mathbf r}_a} = \nabla_a V_b\left[1-\frac{h_b}{3V_b}\frac{\partial V_b}{\partial h_b}\right]^{-1}
     \label{app_3}
     \,,
 \end{equation}
\noindent
where $\nabla_a V_b$ refers to the spatial gradient whereas the grad-h effects are included in the term in brackets.

To estimate $\nabla_a V_b$, the precise form of the volume elements has to be known. A general form for these elements comes from \citet{hop13}, namely:

\begin{equation}
    V_b = \frac{X_b}{k_b}\qquad {\mathrm with}\qquad k_b = \sum_c X_c W_{bc}(h_b)
    \,,
    \label{app_4}
\end{equation}

\noindent 
where $X_b$ is a scalar estimator. Here we have considered two different estimator families, leading to slightly different expressions of the movement and energy equations:

a) {\bf Constant $X_b$}, as for example $X_b= m_b$ and $X_b =1$ both reproducing the standard volume elements $V_b = m_b/\rho_b$. Other choice is $X_b=P^k_b$ where $P$ is the pressure and $k\le 1$. In this case, the estimator is strictly constant only in isobaric systems.

\begin{equation}
  \nabla_a V_b \xrightarrow[b=a]{} \nabla_a V_a = \nabla_a\left(\frac{X_a}{k_a}\right) = -\frac{X_a}{k_a^2}\nabla_a k_a\\ =-\frac{X_a}{k_a^2}\sum_b X_b\nabla_a W_{ab}(h_a)
  \,,
  \label{app_5}
\end{equation}

\begin{equation}
  \nabla_a V_b \xrightarrow[b\ne a]{} \nabla_a V_b = \nabla_a\left(\frac{X_b}{k_b}\right) = -\frac{X_b}{k_b^2}\nabla_a k_b =-\frac{X_b}{k_b^2}~ X_a\nabla_a W_{ab}(h_b)
  \,.
  \label{app_6}
\end{equation}

Combining expressions (\ref{app_3}), (\ref{app_5}), and (\ref{app_6}) with Eq.~(\ref{app_1}), and making use of Eq.~(\ref{stdiad}) to carry out the IA approach of the kernel gradient, the $i$-component of the acceleration of particle $a$ is finally obtained:

\begin{equation}
  \ddot x_{i,a}=-\frac{X_a}{m_a}\sum_b \left[\frac{X_b P_a}{\Omega_a k_a^2} \mathcal A_{i,ab}(h_a) + \frac{X_b P_b}{\Omega_b k_b^2}\mathcal A_{i,ab}(h_b)\right]
  \,,
  \label{app_mom_1} 
\end{equation}

with

\begin{equation}
    \Omega_a= \left[1-\frac{h_a}{3V_a}\frac{\partial V_a}{\partial h_a}\right]=\left[1+\frac{h_a}{3\rho_a}\frac{\partial\rho_a}{\partial h_a}\right]
    \,.
    \label{app_omega_1}
\end{equation}

\vspace{0.2cm}

b) {\bf Adaptive $X_b=m_b/\rho_b^0$},~being $\rho^0_b = \sum_c m_c W_{bc}(h_b)$ the standard SPH density. Because of the normalization condition, this choice leads to a constant \linebreak \mbox{$k_b = \sum_c X_c W_{bc}(h_b)$}. Now we have 

\begin{equation}
\begin{split}
  \nabla_a V_b \xrightarrow[b=a]{} \nabla_a V_a = \nabla_a \left(\frac{X_a}{k_a}\right) = \frac{\nabla_a X_a}{k_a} &= -\frac{m_a}{(\rho^0_a)^2 k_a}\nabla_a\rho^0_a\\
  &=-\frac{m_a}{(\rho^0_a)^2 k_a}\sum_b m_b\nabla_a W_{ab}(h_a)
\,,
\end{split}
  \label{app_7}
\end{equation}

\begin{equation}
\begin{split}
\nabla_a V_b \xrightarrow[b\ne a]{} \nabla_a V_b = \nabla_a \left(\frac{X_b}{k_b}\right) = \frac{\nabla_a X_b}{k_b} &= -\frac{m_b}{(\rho^0_b)^2 k_b}\nabla_a\rho^0_b\\ &=-\frac{m_b m_a}{(\rho^0_b)^2 k_b}\nabla_a W_{ab}(h_b)
\,.
\end{split}
  \label{app_8}
\end{equation}

Combining expressions (\ref{app_3}), (\ref{app_7}), and (\ref{app_8}) with Eq.~(\ref{app_1}), and making use of Eq.~(\ref{stdiad}) to compute the IA approach of the kernel gradient, the $i$-component of the acceleration of particle $a$ is obtained, 

\begin{equation}
  \ddot x_{i,a}=-\sum_b m_b\left[\frac{X_a^2 P_a}{\Omega_a m_a^2~k_a} \mathcal A_{i,ab}(h_a) + \frac{X_b^2 P_b}{\Omega_b m_b^2~k_b}\mathcal A_{i,ab}(h_b)\right] \label{app_mom_2}
  \,.
\end{equation}

According to expression (\ref{app_omega_1}), to estimate $\Omega_a$ it is necessary to know the derivative of the density with respect the smoothing-length $(\partial\rho_a/\partial h_a)$. The result relies in the choice the estimator $X_a$ used to compute the density. 

\begin{itemize}

\item For constant $X_a$,~as for example (but not necessarily), the standard choice $X_a=m_a$:

\begin{equation}
    \frac{\partial\rho_a}{\partial h_a}= \frac{m_a}{X_a}\sum_b X_b \frac{\partial W_{ab}(h_a)}{\partial h_a}
    \,.
    \label{graddens_h_1}
\end{equation}

\item The choice $X_a=(m_a/\rho^0_a)$ requires a bit more algebra:

\begin{equation}
    \frac{\partial\rho_a}{\partial h_a}=\frac{m_a}{X_a}\left(\frac{\partial k_a}{\partial h_a}\right)-\frac{m_a k_a}{X_a^2}\left(\frac{\partial X_a}{\partial h_a}\right)
    \,,
    \label{graddens_h_2}
\end{equation}

with:

\begin{equation}
\begin{split}
 \frac{\partial k_a}{\partial h_a} &= \sum_b\frac{\partial}{\partial h_a} \left(\frac{m_b}{\rho_b^0}\right) W_{ab}(h_a)+ \sum_b X_b \frac{\partial W_{ab}(h_a)}{\partial h_a} \\
 &= -\left(\frac{m_a}{(\rho^0_a)^2}\right) W_{aa}(h_a)\frac{\partial\rho^0_{a}(h_a)}{\partial{h_a}}+\sum_b X_b \frac{\partial W_{ab}(h_a)}{\partial h_a}
 \,,
 \end{split}
 \label{dkdh}
\end{equation}

and, 

\begin{equation}
 \frac{\partial X_a}{\partial h_a}= \frac{\partial}{\partial h_a}\left(\frac{m_a}{\rho^0_a}\right)=-\left(\frac{m_a}{(\rho^0_a)^2}\right)\frac{\partial\rho^0_{a}(h_a)}{\partial{h_a}}=-\left(\frac{m_a}{(\rho^0_a)^2}\right)\sum_b m_b \frac{\partial W_{ab}(h_a)}{\partial h_a}
 \,.
 \label{dXdh}
\end{equation}

Setting expressions (\ref{dkdh}) and (\ref{dXdh}) into Eq.~(\ref{graddens_h_2}) and via the ensuing manipulation, we have: 

\begin{equation}
    \frac{\partial\rho_a}{\partial h_a}=\left[\frac{\rho_a}{\rho^0_a}-X_a W_{aa}(h_a)\right]\sum_b m_b \frac{\partial W_{ab}(h_a)}{\partial h_a}+\frac{m_a}{X_a}\sum_b X_b \frac{\partial W_{ab}(h_a)}{\partial h_a}
    \,.
\end{equation}

\end{itemize}
\end{appendix}
\end{document}